\documentclass[a4paper,10pt,notitlepage,twocolumn]{revtex4-1}
\usepackage[utf8]{inputenc}
\usepackage[ngerman,english]{babel}
\pdfoutput=1
\usepackage{url}
\usepackage{bbm}
\usepackage[T1]{fontenc}
\usepackage[utf8]{inputenc}
\usepackage{graphicx}
\usepackage{eurosym}
\usepackage{amsmath}
\usepackage{amssymb}
\usepackage{hyperref}
\usepackage{amsthm}
\usepackage{subfigure}
\usepackage{setspace}
\usepackage{colortbl}
\usepackage[titletoc,title]{appendix}
\usepackage{natbib}
\usepackage[left=2.0cm,right=2.0cm,top=3.0cm,bottom=3.0cm]{geometry}
\newcommand\numberthis{\addtocounter{equation}{1}\tag{\theequation}}

\begin{document}
 \title{Quantum error correction against photon loss using multi-component cat states}
\author{Marcel Bergmann}
\author{Peter van Loock}
\affiliation{Institut für Physik, Johannes Gutenberg-Universität, Staudingerweg 7, 55128 Mainz, Germany}

\begin{abstract}
\noindent
We analyse a generalised quantum error correction code against
photon loss where a logical qubit is encoded into a subspace of
a single oscillator mode that is spanned by distinct multi-component
cat states (coherent-state superpositions). We present a systematic
code construction that includes the extension of an existing one-photon-loss
code to higher numbers of losses. When subject to a photon loss
(amplitude damping) channel, the encoded qubits are shown to exhibit
a cyclic behaviour where the code and error spaces each correspond to
certain multiples of losses, half of which can be corrected. As another
generalisation we also discuss how to protect logical qudits
against photon losses, and as an application we consider
a one-way quantum communication scheme, in which the encoded
qubits are periodically recovered while the coherent-state
amplitudes are restored as well at regular intervals. 
\end{abstract}
\maketitle
\onecolumngrid
\section{Introduction}

Photons are fundamental carriers of quantum information.
Travelling at the speed of light, they are the optimal choice for
quantum communication. In the form of photonic qubits
each encoded into two orthogonal polarizations, they allow for
a simple way to reach any point on a single-qubit's Bloch sphere
through polarization rotations. Nonetheless, there are also
disadvantages when quantum information is encoded into single
photons. Two-qubit operations acting jointly on two photons
are notoriously hard to achieve and require non-linear interactions.
The biggest drawback for quantum communication, however,
is that photons get quickly absorbed along a communication
channel such as an optical fibre. This prevents the immediate use
of simple photonic qubits for long-distance quantum communication
(with distances of 500-1000 km or more).

There have been several proposals of quantum error correction (QEC)
codes to protect photonic qubits against the effect of photon loss
\cite{bosonic_codes,Banaszek, Glancy}. Some of these codes make use of an extra
number of modes, while each mode contains either zero or one photon
\cite{approximate_codes, Fletcher, QPC}. Other codes do not require a large number of modes,
but instead include the possibility of having more than just one photon in
every mode \cite{bosonic_codes}. Yet other loss codes are intermediate
in terms of mode and photon number \cite{NOON}.
Among all these codes, in general, the total mean photon number
determines the loss scaling of the code: the more photons in the
encoded states are, the more likely some photons will get lost,
i.e., the loss rate goes up. On the other hand, for a fixed mode number,
the inclusion of more photons leads to larger Hilbert (sub)spaces and
the possibility of correcting higher orders of losses.
For codewords with a fixed total photon number $N$, typically
$\sqrt{N} - 1$ photon losses can be exactly corrected \cite{bosonic_codes, QPC, NOON}.
In general, however, there are no simple and efficient schemes
to implement such higher-order photonic loss codes. 

Another approach to optical, loss-adapted QEC is to
encode a logical qubit into a single oscillator mode \cite{GKP,Leghtas, LeghtasPRL}.
Such a code can make explicit use of the infinite-dimensional Hilbert space
already available with just one optical, physical mode. By sticking to a finite-dimensional
(logical qubit) code space, such codes also circumvent existing no-go results
for efficient QEC of logical continuous-variable Gaussian states encoded
into physical, multi-mode Gaussian states \cite{BraunsteinPRL,LloydSlotinePRL,BraunsteinNature, PvLJModOpt, Aoki}
and subject to Gaussian error channels \cite{NisetNogoPRL}, \footnote{
more precisely, non-Gaussian logical states alone, such as logical qubits, are actually not
enough to circumvent those no-go results when both error and recovery channels
are of Gaussian nature \cite{Namiki}. Note that the amplitude-damping (photon-loss)
channel is indeed a Gaussian channel. Nonetheless, non-Gaussian logical states
subject to Gaussian error channels together with non-Gaussian ancilla states \cite{GKP, GlancyKnill}
or with non-Gaussian operations \cite{LeghtasPRL, Leghtas} for the recoveries do the trick.
Similarly, Gaussian logical states subject to non-Gaussian error channels
with only Gaussian recovery operations suffice to circumvent the no-go theorem \cite{PvLJModOpt, Aoki}.}.
The qubit-into-oscillator codes are approximate codes based on non-orthogonal
codewords that become perfect for infinite squeezing \cite{GKP} or for infinitely large coherent-state
amplitudes \cite{LeghtasPRL, Leghtas}. Note that the GKP code with codewords as superpositions of position (quadrature)
eigenstates \cite{GKP} is a universal code, whereas the cat code with codewords
as even cat states (that is superpositions of even photon numbers) \cite{LeghtasPRL,Leghtas} is specifically adapted to
photon loss errors. Recently, also the concept of an approximate single-mode bosonic code
was introduced whose codewords are finite superpositions of certain multiples of the
photon number \cite{newclass}.

The possibility of a generalisation of the one-loss cat code \cite{LeghtasPRL, Leghtas} to higher losses
has been briefly mentioned a couple of times in the literature (see the Conclusions of Ref. \cite{LeghtasPRL} and
Sec.~4.2. on page 15 of Ref. \cite{lifetime}),
including a few more detailed hints about the conceptual character of such an extension in a very recent
publication (Fig.~1 on page 4 and Sec.~V.B. on page 10 of Ref. \cite{newclass}). 
However, as far as we know there is no detailed analysis of a generalised code that includes
a complete and systematic definition of the codewords
as well as a quantitative performance assessment of the code in a full amplitude-damping channel. 
Here we present such an analysis. We will give a very compact definition of the codewords in terms of
eigenvalue equations, expressed in terms of powers of the mode annihilation operators,
for any loss order. This way we will also define the canonical codewords for the respective error
spaces which satisfy the same eigenvalue equations, but differ from the code space-codewords
and the codewords from the other error spaces in their (generalised) number parities.
Thus, a certain instance of the cat code (corresponding to a certain coherent-state amplitude $\alpha$,
a certain loss order $L$, and also a certain logical dimension $d$ for general logical qudits
living in the code and error spaces)
is defined by two sets of eigenvalue equations: one to determine the space (and hence the error syndrome) and another one
to define (together with the former set) the codewords.
We will demonstrate that for the right choice of codewords there is no deformation
of the initial logical qubit (not even for small coherent-state amplitudes, in which case, however,
the codewords begin to overlap significantly). This no-deformation property results
in a rather simple and well structured output density matrix when the encoded state is subject to
a complete loss channel. This feature is also similar to the cyclic behaviour of the one-loss-code
in the simplified photon-annihilation ("photon-jump") error model \cite{LeghtasPRL, Leghtas}, but here extended to
higher losses and for the full, physical amplitude-damping channel. \\

The paper is structured as follows. In Sec.~\ref{sec: cat-code}, we start with a discussion of the
known one-loss code \cite{LeghtasPRL,Leghtas}
and its properties when subject to individual photon loss events as well as
its behaviour in a full loss channel. A generalisation of this code to higher numbers of losses
is presented in Sec.~\ref{sec: generalised cat codes}, again including a discussion of its behaviour
under the two manifestations of photon loss (single loss events and full channel). Coherent-state
superpositions with sufficiently many components may also be utilized to encode logical
quantum information beyond qubits in so-called qudits \cite{holonomic}, and we will determine
the conditions for such qudit cat codes adapted to one and more losses of photons in Sec.~\ref{sec: qudit}.
In Sec.~\ref{sec: oneway communication}, as an example for an application
and in order to illustrate how the loss codes can be
used for a quantum information task, we consider a one-way quantum communication scheme, in which
an encoded qubit is sent along a quantum channel while being recovered at regular intervals
to preserve the qubit information until the end of the channel. The channel here is assumed to
cover a rather long distance (such as 1000 km as a typical distance in long-distance quantum
communication) and in addition to the qubit recoveries, we also show how the decay of the
coherent-state amplitude can be, in principle, dealt with as well for such a large total distance.
After the Conclusions we present several Appendices that include additional technical details
and explanations.

\section{One-loss Cat code}
\label{sec: cat-code}
It is rather well-known that there is a two-fold effect when a cat state, i.e., a superposition of two distinct coherent states
such as $\propto |\alpha\rangle+|-\alpha\rangle$ is subject to a full photon loss (amplitude damping) channel.
On the one hand, the coherent-state amplitude $\alpha$ in each term is attenuated depending on the channel transmission parameter,
$\sqrt{\gamma}\alpha$, corresponding to an exponential amplitude decay with distance. On the other hand, a random phase flip occurs that
incoherently mixes the initial cat state with its phase-flipped version such as $\propto |\alpha\rangle-|-\alpha\rangle$, where the flip
probability also depends on the channel transmission $\gamma$ and on the initial amplitude $\alpha$. In a cat-state qubit encoding \cite{Ralph,RalphLundPRL},
a loss-induced phase flip of a logical qubit could be corrected when the qubit is encoded into an additional layer of a multi-qubit
repetition code composed of three or more logical cat qubits (i.e., by adding two or more physical oscillator modes) \cite{Glancy, Wickert1, Wickert2}.
A conceptually more innovative approach, however, would stick to a single oscillator mode and instead exploit more than just two
(near-)orthogonal coherent-state components of that mode
(i.e., exploiting a manifold with dimension larger than two in the oscillator's phase space).
While it is obvious that this approach enables one to reach higher dimensions, it is not immediately clear how this can provide protection
against photon losses. In Ref.~\cite{LeghtasPRL,Leghtas}, however, it was shown that by constructing two (near-)orthogonal codewords both in the form of
even cat states (those with only even photon-number terms) a logical qubit can be encoded that remains intact under the effect
of a lost photon, as the qubit is then mapped onto an orthogonal error space that is spanned by two (near-)orthogonal codewords both in the form of
odd cat states (those with only odd photon-number terms).

Formally, for the even cat code given in \cite{LeghtasPRL,Leghtas}, the basic codewords are certain +1 eigenstates of the number parity operator
$(-1)^{\hat{n}}$:

\begin{equation}
\label{eqn: catcode} 
\begin{aligned}
 |\bar{0}_{+}\rangle&=\frac{1}{\sqrt{N_{+}}}(|\alpha\rangle+|-\alpha\rangle),\\
 |\bar{1}_{+}\rangle&=\frac{1}{\sqrt{N_{+}}}(|i\alpha\rangle+|-i\alpha\rangle),\\
\end{aligned}
\end{equation}
 with normalisation constant $N_{\pm}=2\pm 2\exp(-2\alpha^{2})$ ($N_{-}$ for later). Throughout we assume $\alpha \in \mathbb{R}$. By writing the coherent 
 states in the Fock basis, one can easily confirm that both codewords have only even photon number terms, 
 \begin{equation}
\label{eqn: evenfock} 
\begin{aligned}
 |\bar{0}_{+}\rangle&=\frac{2e^{-\alpha^{2}/2}}{\sqrt{N_{+}}}\left(|0\rangle+\frac{\alpha^{2}}{\sqrt{2}}|2\rangle+\frac{\alpha^{4}}{2\sqrt{6}}|4\rangle+...\right),\\
 |\bar{1}_{+}\rangle&=\frac{2e^{-\alpha^{2}/2}}{\sqrt{N_{+}}}\left(|0\rangle-\frac{\alpha^{2}}{\sqrt{2}}|2\rangle+\frac{\alpha^{4}}{2\sqrt{6}}|4\rangle-...\right).\\
\end{aligned}
\end{equation}
 These two so-called even cat states are, in general, not orthogonal, but for large $\alpha$, as $e^{-\alpha^{2}/2}\alpha^{k}\rightarrow 0$, an infinite superposition
 of nearly equally weighted even number states is obtained for each codeword, $|\bar{0}_{+}\rangle \propto |0\rangle+|2\rangle+|4\rangle+... $
 and $|\bar{1}_{+}\rangle \propto |0\rangle-|2\rangle+|4\rangle-... $ and thus $\langle\bar{0}_{+}|\bar{1}_{+}\rangle\approx 0$ (notice the alternating sign in
 $|\bar{1}_{+}\rangle$).
 For general $\alpha$, their overlap is (see App. \ref{sec: oneloss})
 \begin{equation}
 \label{eq: plusoverlap}
 \begin{aligned}
\langle\bar{0}_{+}|\bar{1}_{+}\rangle= &\frac{1}{N_{+}}\left(\langle \alpha|i\alpha\rangle+\langle \alpha|-i\alpha\rangle+\langle -\alpha|i\alpha\rangle+\langle -\alpha|-i\alpha\rangle\right)\\
&=\frac{\cos(\alpha^{2})}{\cosh(\alpha^{2})},
 \end{aligned}
 \end{equation}
 which indeed goes to zero in the limit $\alpha\rightarrow \infty$. 
Instead of the codewords in Eq. \eqref{eqn: catcode}, as an alternative qubit basis, we may also use the two \textit{orthogonal} states
\begin{equation}
\label{eq: xbasis}
 |\bar{0}_{+}\pm \bar{1}_{+}\rangle=\frac{1}{\sqrt{N_{\pm}^{\prime}}}(|\alpha\rangle+|-\alpha\rangle\pm |i\alpha\rangle\pm |-i\alpha\rangle),
\end{equation}
which span the same (even) code space as $\{|\bar{0}_{+}\rangle, |\bar{1}_{+}\rangle\}$ do and hence represent the same (even) cat code 
($N_{\pm}^{\prime}$ are some normalisation constants). Their exact orthogonality (for any $\alpha$) can be immediately seen in the Fock basis:
\begin{equation}
 \begin{aligned}
  |\bar{0}_{+}+ \bar{1}_{+}\rangle&=\frac{4e^{-\alpha^{2}/2}}{\sqrt{N_{+}^{\prime}}}\left(|0\rangle+\frac{\alpha^{4}}{2\sqrt{6}}|4\rangle+\frac{\alpha^{8}}{24\sqrt{70}}|8\rangle+...\right),\\
 |\bar{0}_{+}- \bar{1}_{+}\rangle&=\frac{4e^{-\alpha^{2}/2}}{\sqrt{N_{-}^{\prime}}}\left(\frac{\alpha^{2}}{\sqrt{2}}|2\rangle+\frac{\alpha^{6}}{12\sqrt{5}}|6\rangle+\frac{\alpha^{10}}{720\sqrt{7}}|10\rangle+...\right).
 \end{aligned}
\end{equation}

Here we refer to the non-orthogonal codewords  $|\bar{0}_{+}\rangle$ and $|\bar{1}_{+}\rangle$ as the (approximate) logical Pauli-$\bar{Z}$ basis, and in 
this sense, the states $|\bar{0}_{+}\pm \bar{1}_{+}\rangle$ can be thought of a logical Pauli-$\bar{X}$ basis obtained by taking an equally weighted sum or 
difference of the two $\bar{Z}$ eigenstates. This is similar to the cat-qubit encoding of Ref. \cite{RalphLundPRL} when two non-orthogonal phase-rotated coherent
states $\{|\pm \alpha\rangle\}$ form the computational $\bar{Z}$ basis, while the two orthogonal even and odd cat states 
$\{|\alpha\rangle \pm |-\alpha\rangle\}$ correspond to the Hadamard-transformed, logical $\bar{X}$ basis (this encoding, however, does not represent
a loss code that allows to correct a certain non-zero number of photon losses and it corresponds to the 0th order of our family of generalised cat codes, 
see next section).\\
Although $\{|\bar{0}_{+}\rangle, |\bar{1}_{+}\rangle\}$ and $\{|\bar{0}_{+}\pm \bar{1}_{+}\rangle\}$ represent the same code, we will see that, nonetheless,
the choice of codewords, for example the $\bar{Z}$ or $\bar{X}$ basis, does make a difference when assessing the code's performance in a physical loss channel.
This is related to the fact that the code is an approximate code, for which there is not a clear distinction between correctable errors (exactly satisfying
the Knill-Laflamme (KL) conditions \cite{N-C}, see App. \ref{sec: QEC}) and uncorrectable errors (violating the KL conditions) like for an exact code. For the approximate
cat code, those errors that are, in principle, correctable may still give violations of the KL conditions, however, these violations go away in the limit of large 
amplitudes $\alpha$. For general $\alpha$ values, it then depends on the choice of codewords what particular KL conditions are violated and, as a result, 
what particular logical  errors occur. These logical errors reduce the (input-state-dependent qubit) fidelity, which is further reduced by the uncorrectable errors
(which remain uncorrectable even when $\alpha\rightarrow \infty$ and which occur more frequently when $\alpha$ is large, see below).\\
As one type of violation of the KL conditions can be avoided at least in the 0th order (i.e., the orthogonality condition of the initial codewords) 
for the basis $\{|\bar{0}_{+}\pm \bar{1}_{+}\rangle\}$, independently of $\alpha$, it appears beneficial to choose this basis. However, for finite $\alpha$, these codewords lead to a deformation
of the logical qubit, i.e., the norms of the codewords after an otherwise correctable error (such as a one-loss-error for the one-loss-code) change depending on the
specific codeword. This latter effect of qubit deformation turns out to be highly undesirable when the full photon loss channel is considered and so our choice
of codewords will be the non-orthogonal $\{|\bar{0}_{+}\rangle, |\bar{1}_{+}\rangle\}$-basis. These codewords do not lead to a qubit deformation, i.e., 
the change in the norm of either codeword after a one-loss-error (or any other correctable error such as 0,4,8,12,... or 5,9,13,... losses of photons, see below) is independent
of the codeword for any $\alpha$. This no-deformation property of the codewords means that the nice cyclicity feature of the cat code for a simplified, unphysical photon-loss
error model, as we discuss next, can be effectively taken over to the physical model of a full loss channel. The only remaining effects that have to be dealt
with then come from the non-orthogonality of the codewords  $|\bar{0}_{+}\rangle$ and  $|\bar{1}_{+}\rangle$ before and after an error (i.e., 
in the code and the error spaces, as it becomes manifest through violations of the corresponding KL conditions).
In App. \ref{sec: oneloss}, we present a detailed discussion of the KL criteria for the various error models.\\

In order to understand the behaviour of the codewords under photon loss, it is conceptually useful to first model the 
effects of the channel by individual photon loss and simply apply the annihilation operator $\hat{a}$ to the codewords. Higher losses
are analogously represented by higher powers of $\hat{a}$. It also turns out to be advantageous to look at even and odd powers
 separately:
 \begin{equation}
 \begin{aligned}
  \hat{a}^{2k}|\bar{0}_{+}\rangle&=\alpha^{2k}\frac{1}{\sqrt{N_{+}}}(|\alpha\rangle+|-\alpha\rangle),\\
  \hat{a}^{2k}|\bar{1}_{+}\rangle&=(-1)^{k}\alpha^{2k}\frac{1}{\sqrt{N_{+}}}(|i\alpha\rangle+|-i\alpha\rangle),\\
  \hat{a}^{2k+1}|\bar{0}_{+}\rangle&=\alpha^{2k+1}\frac{1}{\sqrt{N_{+}}}(|\alpha\rangle-|-\alpha\rangle),\\
  \hat{a}^{2k+1}|\bar{1}_{+}\rangle&=i(-1)^{k}\alpha^{2k+1}\frac{1}{\sqrt{N_{+}}}(|i\alpha\rangle-|-i\alpha\rangle),\\
  \end{aligned}
  \end{equation}
where $k=0,1,2,...$. According to this simplified loss model, a logical qubit of the (unnormalised) form $|\bar{\psi}\rangle=a|\bar{0}_{+}\rangle+b|\bar{1}_{+}\rangle$
 evolves cyclically into the following four (unnormalised) states \cite{Leghtas, LeghtasPRL},
 \begin{equation}
 \begin{aligned}
  |\bar{\psi}\rangle_{4k}&=a|\bar{0}_{+}\rangle+b|\bar{1}_{+}\rangle,\\
  |\bar{\psi}\rangle_{4k+1}&=a|\bar{0}_{-}\rangle+ib|\bar{1}_{-}\rangle,\\
  |\bar{\psi}\rangle_{4k+2}&=a|\bar{0}_{+}\rangle-b|\bar{1}_{+}\rangle,\\
  |\bar{\psi}\rangle_{4k+3}&=a|\bar{0}_{-}\rangle-ib|\bar{1}_{-}\rangle,
 \end{aligned}
 \end{equation}
 depending on whether the number of lost photons is 0,4,8,... or 1,5,9,... or 2,6,10,... or 3,7,11,..., respectively.
 Here, we defined the non-orthogonal basic codewords for the error space as
 \begin{equation}
 \begin{aligned}
 |\bar{0}_{-}\rangle&=\frac{1}{\sqrt{N_{-}}}(|\alpha\rangle-|-\alpha\rangle),\\
 |\bar{1}_{-}\rangle&=\frac{1}{\sqrt{N_{-}}}(|i\alpha\rangle-|-i\alpha\rangle),\\
\end{aligned}
 \end{equation}
which are two so-called odd cat states with only odd photon number terms,
 \begin{equation}
\label{eqn: oddfock} 
\begin{aligned}
 |\bar{0}_{-}\rangle&=\frac{2e^{-\alpha^{2}/2}\alpha}{\sqrt{N_{-}}}\left(|1\rangle+\frac{\alpha^{2}}{\sqrt{6}}|3\rangle+\frac{\alpha^{4}}{2\sqrt{30}}|5\rangle+...\right),\\
 |\bar{1}_{-}\rangle&=\frac{2e^{-\alpha^{2}/2}i\alpha}{\sqrt{N_{-}}}\left(|1\rangle-\frac{\alpha^{2}}{\sqrt{6}}|3\rangle+\frac{\alpha^{4}}{2\sqrt{30}}|5\rangle-...\right).\\
\end{aligned}
\end{equation}
Again, these two codewords approach an orthogonal qubit basis, this time in the odd-parity error space, when $\alpha$
is sufficiently large (notice the alternating sign in $|\bar{1}_{-}\rangle$ inherited from $|\bar{1}_{+}\rangle$).
In fact, the overlap between $|\bar{0}_{-}\rangle$ and $|\bar{1}_{-}\rangle$ is
 \begin{equation}
 \label{eq: minusoverlap}
 \begin{aligned}
\langle\bar{0}_{-}|\bar{1}_{-}\rangle=&\frac{1}{N_{-}}\left(\langle \alpha|i\alpha\rangle-\langle \alpha|-i\alpha\rangle-\langle -\alpha|i\alpha\rangle+\langle -\alpha|-i\alpha\rangle\right)\\
&=\frac{i\sin(\alpha^{2})}{\sinh(\alpha^{2})},
 \end{aligned}
 \end{equation}
which again can be made arbitrarily small by increasing $\alpha$.
The code and error spaces can be characterised by their photon number parity (even/odd) and thus
are perfectly distinguishable. However, there can be uncorrectable phase-flip errors of the logical qubit when it is mapped back to the even code space after
half a cycle, $|\bar{\psi}\rangle_{4k+2}=a|\bar{0}_{+}\rangle-b|\bar{1}_{+}\rangle$, or when it is mapped again onto the odd error space before the end of a cycle,
$|\bar{\psi}\rangle_{4k+3}=a|\bar{0}_{-}\rangle-ib|\bar{1}_{-}\rangle$. Otherwise the qubit remains intact either in the code space,  
$|\bar{\psi}\rangle_{4k}=a|\bar{0}_{+}\rangle+b|\bar{1}_{+}\rangle$, or in the error space,
$|\bar{\psi}\rangle_{4k+1}=a|\bar{0}_{-}\rangle+ib|\bar{1}_{-}\rangle$ (in which case it is transformed by a known and fixed phase gate). Once the parity is detected,
the qubit is recovered and no further correction step is needed. The uncorrectable errors lead to a non-unit fidelity, when the
actual physical loss channel is considered which we do next.\\ 
Photon loss, for example occurring in an optical fibre, 
is described by the amplitude damping (AD) channel \cite{bosonic_codes}. In the single-mode AD model, the loss
of exactly $k$ photons can be expressed by a non-unitary error operator,
\begin{equation}
A_{k}=\sum\limits_{n=k}^{\infty}\sqrt{\binom{n}{k}}\sqrt{\gamma}^{n-k}\sqrt{1-\gamma}^{k}|n-k\rangle\langle n|=\sqrt{\frac{(1-\gamma)^{k}}{k!}}\sqrt{\gamma}^{\hat{n}}\hat{a}^{k},
\end{equation}
 $\forall k\in \{0,1,\cdots,\infty\}$ and with the number operator $\hat{n}=\hat{a}^{\dagger}\hat{a}$ in $\sqrt{\gamma}^{\hat{n}}$ which describes the amplitude decay. The probability of losing one photon is $1-\gamma$, which is, for instance, 
 related to the length of the path the photon travels through an optical fibre 
 \footnote{In an optical fibre, one has $\gamma=\exp\left(-\frac{l}{L_{att}}\right)$, where $l$ is the propagation distance of the photon in the optical fibre and $L_{att}=22~\text{km}$ is
 the attenuation length.}. Furthermore, note that
 $A_{k}\geq 0$ and $\sum\limits_{k=0}^{\infty}A_{k}^{\dagger}A_{k}=\mathbbm{1}$. 
 The action of AD on an arbitrary input state $\rho$ is
 \begin{equation}\label{AD}
\rho\rightarrow\rho_{f}=\sum\limits_{k=0}^{\infty}A_{k}\rho A_{k}^{\dagger}.
\end{equation}
The full loss channel is now a (complete-positive) trace-preserving map that incorporates all possible individual photon loss events as well as the effect
of amplitude decay.\\
Let us now study the action of AD on the encoding in Eq. \eqref{eqn: catcode}. The somewhat lengthy
calculations are presented in App. \ref{sec: oneloss} and the channel evolution of a normalised logical qubit $|\bar{\psi}\rangle$ is found to be
 \begin{equation}
 \label{eq: output}
 \begin{aligned}
 \bar{\rho}&=\widetilde{p}_{0} \left(\frac{a|\widetilde{0}_{+}\rangle+b|\widetilde{1}_{+}\rangle}{\sqrt{1+2\operatorname{Re}(ab^{*})\frac{\cos(\gamma \alpha^{2})}{\cosh(\gamma \alpha^{2})}}}\right)\times H.c.\\
 &+\widetilde{p}_{1} \left(\frac{a|\widetilde{0}_{-}\rangle+ib|\widetilde{1}_{-}\rangle}{\sqrt{1-2\operatorname{Re}(ab^{*})\frac{\sin(\gamma\alpha^{2})}{\sinh(\gamma\alpha^{2})}}}\right)\times H.c.\\
 &+\widetilde{p}_{2} \left(\frac{a|\widetilde{0}_{+}\rangle-b|\widetilde{1}_{+}\rangle}{\sqrt{1-2\operatorname{Re}(ab^{*})\frac{\cos(\gamma\alpha^{2})}{\cosh(\gamma\alpha^{2})}}}\right)\times H.c.\\
 &+\widetilde{p}_{3} \left(\frac{a|\widetilde{0}_{-}\rangle-ib|\widetilde{1}_{-}\rangle}{\sqrt{1+2\operatorname{Re}(ab^{*})\frac{\sin(\gamma\alpha^{2})}{\sinh(\gamma\alpha^{2})})}}\right)\times H.c.~~ .\\
 \end{aligned}
\end{equation}
 The statistical weights in this mixture are given by
 \begin{equation}
 \label{eq: L1prob}
 \begin{aligned}
 \widetilde{p_{0}}&=\frac{1+2\operatorname{Re}(ab^{*})\frac{\cos(\gamma\alpha^{2})}{\cosh(\gamma\alpha^{2})}}{1+2\operatorname{Re}(ab^{*})\frac{\cos(\alpha^{2})}{\cosh(\alpha^{2})}}p_{0},\\
 \widetilde{p_{1}}&=\frac{1-2\operatorname{Re}(ab^{*})\frac{\sin(\gamma\alpha^{2})}{\sinh(\gamma\alpha^{2})}}{1+2\operatorname{Re}(ab^{*})\frac{\cos(\alpha^{2})}{\cosh(\alpha^{2})}}p_{1},\\
 \widetilde{p_{2}}&=\frac{1-2\operatorname{Re}(ab^{*})\frac{\cos(\gamma\alpha^{2})}{\cosh(\gamma\alpha^{2})}}{1+2\operatorname{Re}(ab^{*})\frac{\cos(\alpha^{2})}{\cosh(\alpha^{2})}}p_{2},\\
 \widetilde{p_{3}}&=\frac{1+2\operatorname{Re}(ab^{*})\frac{\sin(\gamma\alpha^{2})}{\sinh(\gamma\alpha^{2})}}{1+2\operatorname{Re}(ab^{*})\frac{\cos(\alpha^{2})}{\cosh(\alpha^{2})}}p_{3},\\
 \end{aligned}
 \end{equation}
 
 where
 \begin{equation}
 \begin{aligned}
 p_{0}&=\frac{\cosh(\gamma \alpha^{2})}{2\cosh(\alpha^{2})}\left(\cos[-\alpha^{2}(1-\gamma)]+\cosh[-\alpha^{2}(1-\gamma)]\right),\\
 p_{1}&=\frac{\sinh(\gamma \alpha^{2})}{2\cosh(\alpha^{2})}\left(\sin[-\alpha^{2}(1-\gamma)]-\sinh[-\alpha^{2}(1-\gamma)]\right),\\
 p_{2}&=\frac{\cosh(\gamma \alpha^{2})}{2\cosh(\alpha^{2})}\left(-\cos[-\alpha^{2}(1-\gamma)]+\cosh[-\alpha^{2}(1-\gamma)]\right),\\
 p_{3}&=\frac{\sinh(\gamma \alpha^{2})}{2\cosh(\alpha^{2})}\left(-\sin[-\alpha^{2}(1-\gamma)]+\sinh[-\alpha^{2}(1-\gamma)]\right),
 \end{aligned}
 \end{equation}

  \begin{figure}[t!]
\centering
\includegraphics[width=0.6\textwidth]{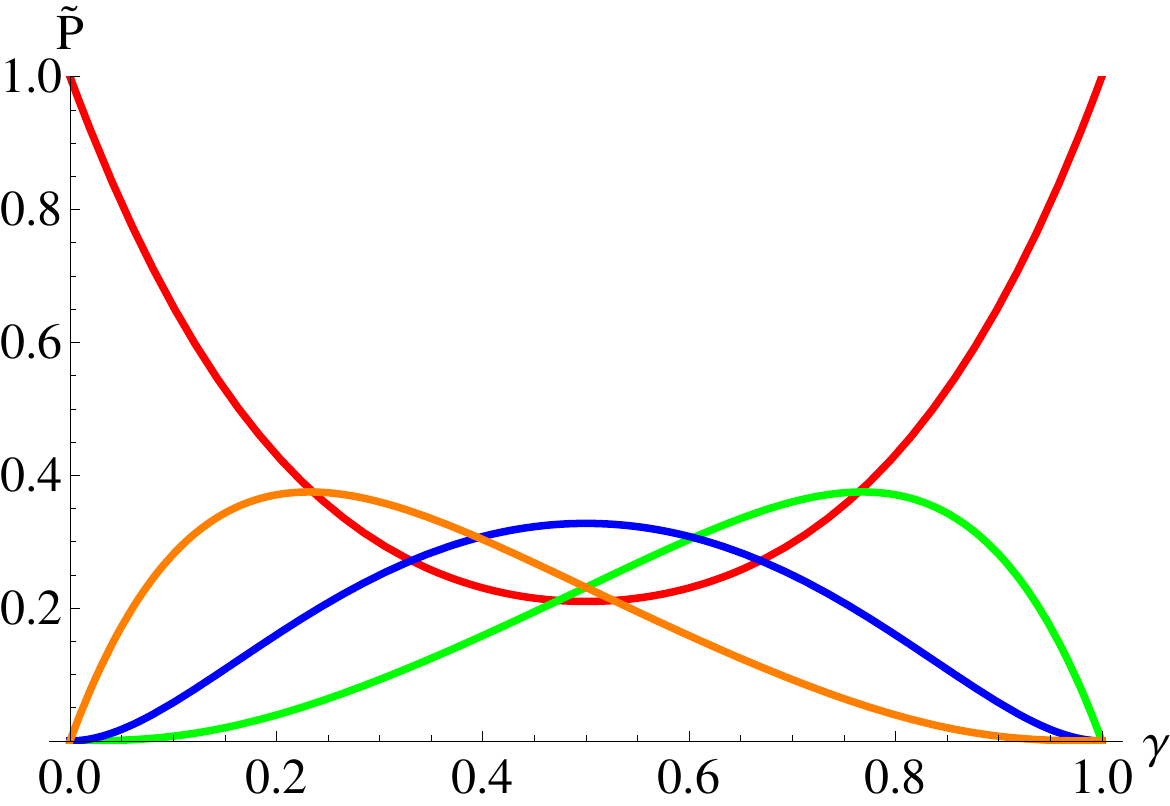}
\caption{
Statistical weights for $\alpha=2$ and $a=b=\frac{1}{\sqrt{2}}$: $\widetilde{p_{0}}$ (red), $\widetilde{p_{1}}$ (green),
$\widetilde{p_{2}}$ (blue), $\widetilde{p_{3}}$ (orange) as a function of the damping parameter $\gamma$ (no damping means $\gamma=1$). Thus, the red and the green curves represent the weights of the correctable errors (0,4,8,... and 
1,5,9,... losses), while the blue and orange curves correspond to the uncorrectable errors (2,6,10,... and 3,7,11,... losses). Note that the $a,b$-dependence can lead to a different qualitative behaviour
of the probabilities for different logical qubits.}
\label{fig: legprob}
\end{figure}
\begin{figure}[t!]
\centering
\includegraphics[width=0.6\textwidth]{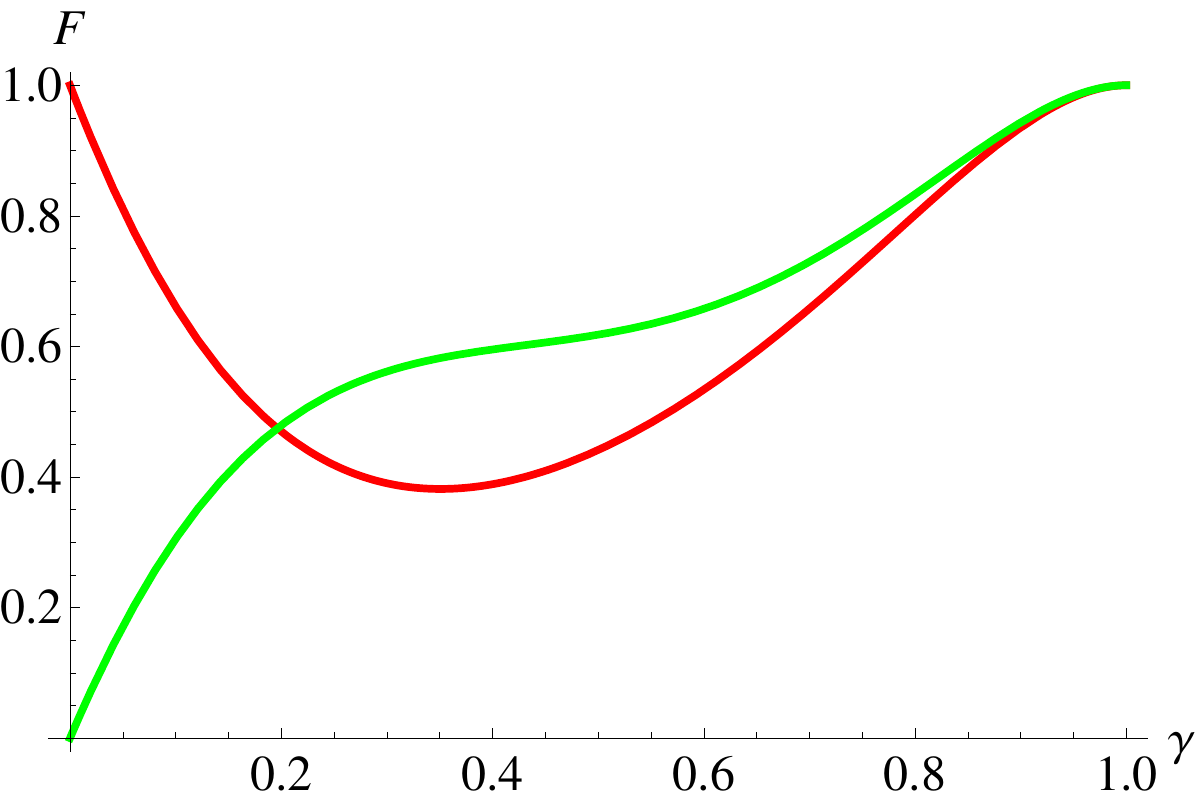}
\caption{Bound $F$ on  worst-case fidelities with a logical qubit $a=b=\frac{1}{\sqrt{2}}$ (red) and $a=-b=\frac{1}{\sqrt{2}}$ (green) for $\alpha=2$
as a function of the damping parameter $\gamma$ (no damping means $\gamma=1$). The actual lower bound on $F_{wc}$ is given by the minimum of the 
two curves for each $\gamma$.}
\label{fig: legfid}
\end{figure}

\noindent
 are the loss probabilities for the individual codewords. The states in the mixture above are damped compared to the input states ($\alpha\rightarrow \sqrt{\gamma}\alpha$),
 which is denoted by the transition $|\bar{0}_{+}\rangle\rightarrow|\widetilde{0}_{+}\rangle$ etc. throughout. Although the complex coefficients of
 the logical input qubit state are normalized as usual, $|a|^{2}+|b|^{2}=1$, note that because of the finite overlap between the codewords in the code and error spaces, an extra factor depending on the input qubit state occurs in the probabilities. 
 This is also related to the fact that the encoding is not an exact quantum error correction code, but only an approximate one (see App. \ref{sec: QEC}).
 The channel output state $\bar{\rho}$ in Eq. \eqref{eq: output} still reflects the cyclic behaviour of the code under individual photon loss events owing to the
 use of the $\bar{Z}$-basis codewords for the logical qubit (thus, avoiding its deformation and a resulting mixture of infinitely 
 many deformed qubits corresponding to infinitely many different loss events). The choice of the logical basis becomes irrelevant only when 
 $\alpha\rightarrow \infty$. Besides the damping of $\alpha$, an uncorrectable phase flip occurs whenever $2,6,10,...$ or $3,7,11,...$ photons are lost. Any other loss errors belong to the correctable set.\\ 
 The error correction works by a QND-type parity measurement which distinguishes between even and odd photon numbers. For this encoding,
 we define an input-state-dependent fidelity as the sum of the statistical weights of the correctable components in the mixture,
 \begin{equation}
 F(a,b)= \widetilde{p_{0}}(a,b)+\widetilde{p_{1}}(a,b). 
 \end{equation}
 The worst-case fidelity $F_{wc}$ is then lower-bounded as
 
 \begin{equation}
 \label{eq: bound}
 F_{wc}\geq  \min_{a,b} F(a,b)\equiv F. 
 \end{equation}
 \noindent
 This bound $F$ is the minimum of the probabilities of correctable errors over all input states. How this lower bound can be understood is explained
 in App. \ref{sec: fidelity}.\\ 
The fidelity $F$ for the one-loss cat code (for a balanced logical qubit minimizing $F(a,b)$, see App. \ref{sec: fidelity})  is shown in Fig. \ref{fig: legfid}. 
The actual $F_{wc}$ is at least as large as plotted there, so that the minimal performance can be inferred. The statistical weights from Eq. \eqref{eq: L1prob} are shown in
Fig. \ref{fig: legprob}, also for a balanced logical qubit.

\section{Generalised Cat codes}
\label{sec: generalised cat codes}
Let us now generalize the one-loss code to include higher losses and state the defining equations for the codewords $|\bar{0}\rangle$ and
$|\bar{1}\rangle$ of an approximate qubit QECC that is capable of correcting $L$ losses:

\begin{equation}
\label{eq: definition}
\begin{aligned}
 \exp\left(\frac{2\pi i \hat{n}}{L+1}\right)|\bar{0}\rangle&=|\bar{0}\rangle,\\
 \exp\left(\frac{2\pi i \hat{n}}{L+1}\right)|\bar{1}\rangle&=|\bar{1}\rangle,\\
 (\hat{a}^{L+1}-\alpha^{L+1})|\bar{0}\rangle&=0,\\
 (\hat{a}^{L+1}+\alpha^{L+1})|\bar{1}\rangle&=0.\\
\end{aligned}
\end{equation}
Here, $\hat{n}=\hat{a}^{\dagger}\hat{a}$ is again the number operator. We will refer to the first two equations as the 'parity conditions' that determine the error
syndrome and hence the subspace in which the qubit resides after an error occurred (one code space and $L$ error spaces).
The current choice of eigenvalue +1 for the two parity conditions in Eq. \eqref{eq: definition} corresponds to (the codewords of) the original 
code space. The error spaces spanned by two codewords with another parity are described by parity conditions with other phase factors as eigenvalues,
see below and App. \ref{sec: derivation}. This is reminiscent of the stabilizer formalism for QEC in terms of Pauli operators \cite{Gottesman}. The last two equations in Eq. \eqref{eq: definition} define 
the codewords in every subspace and remain unchanged for different parities (subspaces), i.e., both codewords are always zero-eigenstates of the
corresponding (generally nonlinear) expressions for the mode operator $\hat{a}$.\\
As shown in App. \ref{sec: derivation} , the (unnormalised) solutions for general $L$ can be written as superpositions of coherent states\newpage
\begin{equation}
\begin{aligned}
 |\bar{0}\rangle&=\sum\limits_{k=0}^{L}|\alpha \exp\left(\frac{2k\pi i }{L+1}\right)\rangle,\\
 |\bar{1}\rangle&=\sum\limits_{k=1}^{L+1}|\alpha \exp\left(\frac{(2k-1)\pi i }{L+1}\right)\rangle.\\
 \end{aligned}
 \end{equation}
 For $L=0$, one obtains the coherent-state encoding $|\bar{0}\rangle=|\alpha\rangle$ and $|\bar{1}\rangle=|-\alpha\rangle$
 presented in Refs. \cite{RalphLundPRL,Ralph} that provides no intrinsic loss protection.
 The $L=1$ case corresponds to the one-loss cat code reviewed in Section \ref{sec: cat-code}. Setting $L=2$ corresponds to a two-loss code,
 for which the unnormalised codewords become
 \begin{equation}
 \begin{aligned}
  |\bar{0}\rangle&=|\alpha\rangle+|\alpha \exp\left(\frac{2\pi i}{3}\right)\rangle+|\alpha \exp\left(-\frac{2\pi i}{3}\right)\rangle,\\
 |\bar{1}\rangle&=|\alpha \exp\left(\frac{\pi i}{3}\right)\rangle+|\alpha \exp\left(\pi i\right)\rangle+|\alpha \exp\left(-\frac{\pi i}{3}\right)\rangle.\\
 \end{aligned}
 \end{equation}
 These are both superpositions of number terms of multiples of three (see App. \ref{sec: twoloss}).
As can easily be checked using the defining equations in Eq. \eqref{eq: definition}, a logical qubit $|\bar{\psi}\rangle=a|\bar{0}\rangle+b|\bar{1}\rangle$
then evolves cyclically under the simplified error model (similar to Sec. \ref{sec: cat-code}) as 
\begin{equation}
\label{eq: L2general}
\begin{aligned}
 \hat{a}^{3k} |\bar{0}\rangle&=\alpha^{3k}|\bar{0}\rangle,\\
 \hat{a}^{3k} |\bar{1}\rangle&=(-1)^{k}\alpha^{3k}|\bar{1}\rangle,\\
 \hat{a}^{3k+1}|\bar{0}\rangle&=\alpha^{3k+1}(|\alpha\rangle+\exp\left(\frac{2\pi i}{3}\right) |\alpha\exp\left(\frac{2\pi i}{3}\right)\rangle
 +\exp\left(-\frac{2\pi i}{3}\right) |\alpha\exp\left(-\frac{2\pi i}{3}\right)\rangle),\\
 \hat{a}^{3k+1}|\bar{1}\rangle&=(-1)^{k}\alpha^{3k+1}\exp\left(\frac{\pi i}{3}\right)(|\alpha\exp\left(\frac{\pi i}{3}\right)\rangle+\exp\left(\frac{2\pi i}{3}\right) |\alpha\exp\left(\pi i\right)\rangle
 +\exp\left(-\frac{2\pi i}{3}\right) |\alpha\exp\left(-\frac{\pi i}{3}\right)\rangle),\\
 \hat{a}^{3k+2}|\bar{0}\rangle&=\alpha^{3k+2}(|\alpha\rangle+\exp\left(-\frac{2\pi i}{3}\right) |\alpha\exp\left(\frac{2\pi i}{3}\right)\rangle
 +\exp\left(\frac{2\pi i}{3}\right) |\alpha\exp\left(-\frac{2\pi i}{3}\right)\rangle),\\
  \hat{a}^{3k+2}|\bar{1}\rangle&=(-1)^{k}\alpha^{3k+2}\exp\left(\frac{2\pi i}{3}\right)(|\alpha\exp\left(\frac{\pi i}{3}\right)\rangle+\exp\left(-\frac{2\pi i}{3}\right) |\alpha\exp\left(\pi i\right)\rangle
 +\exp\left(\frac{2\pi i}{3}\right) |\alpha\exp\left(-\frac{\pi i}{3}\right)\rangle),\\
\end{aligned}
\end{equation}
where again $k=0,1,2,...$. Similar to $L=1$, we encounter a cyclic behaviour. For even $k$ (especially $k=0$ corresponding 
to 0, 1 and 2 losses), there are no $k$-dependent phase flips (the factors $(-1)^{k}$ in front of the transformed $|\bar{1}\rangle$-codewords
in lines 2, 4 and 6 on the rhs of Eq. \eqref{eq: L2general}, see also below) and only fixed, $k$-independent phase factors (in front of the transformed $|\bar{1}\rangle$-codewords). 
The parities for the one- and two-loss cases change compared to the zero-loss case from 0,3,6,... 
to 2,5,8,... and 1,4,7,..., respectively (see App. \ref{sec: twoloss}).\\

\begin{figure}[t!]
  \centering
  \subfigure[$L=0$]{\includegraphics[width=0.23\textwidth]{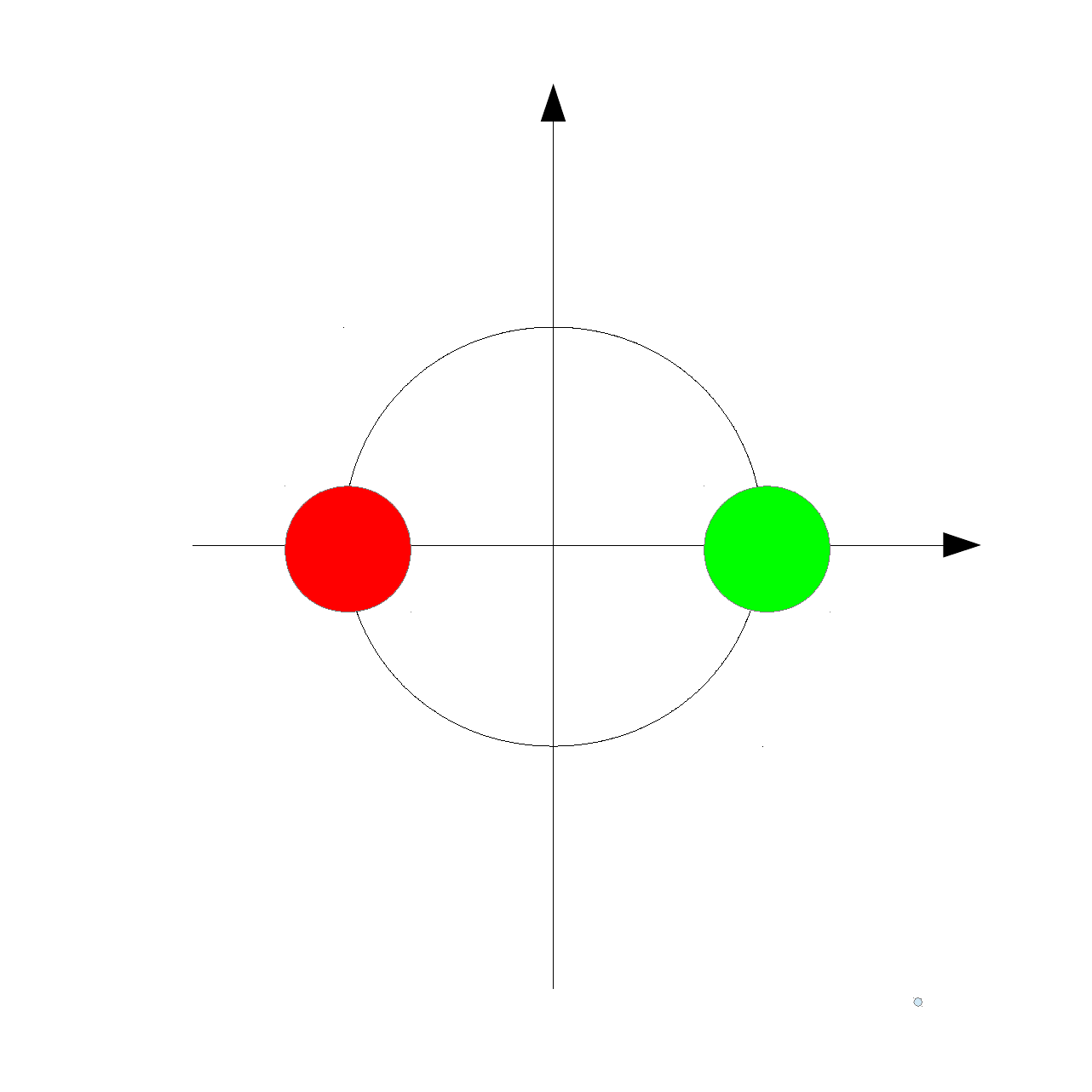}}\quad
  \subfigure[$L=1$]{\includegraphics[width=0.23\textwidth]{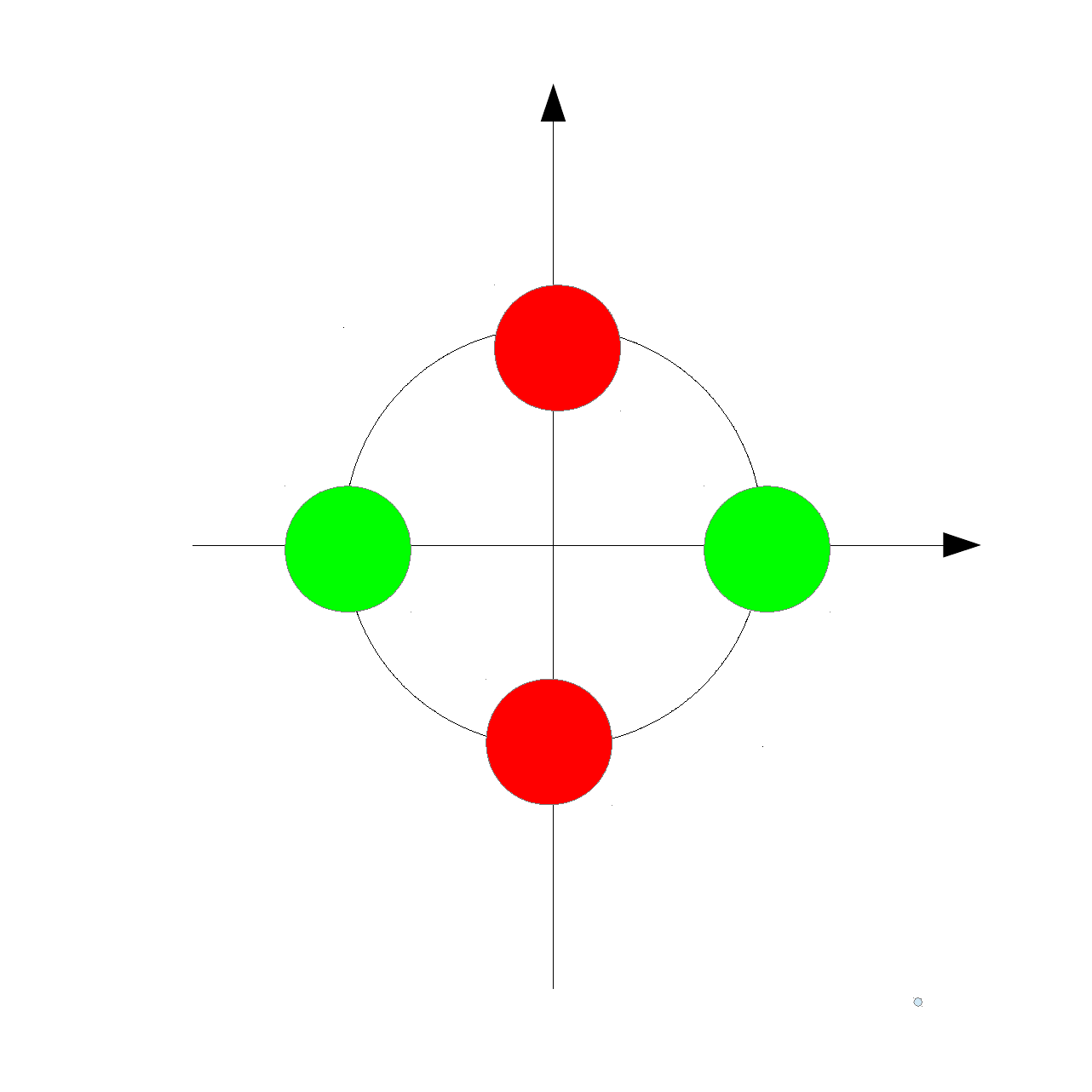}}\quad
  \subfigure[$L=2$]{\includegraphics[width=0.23\textwidth]{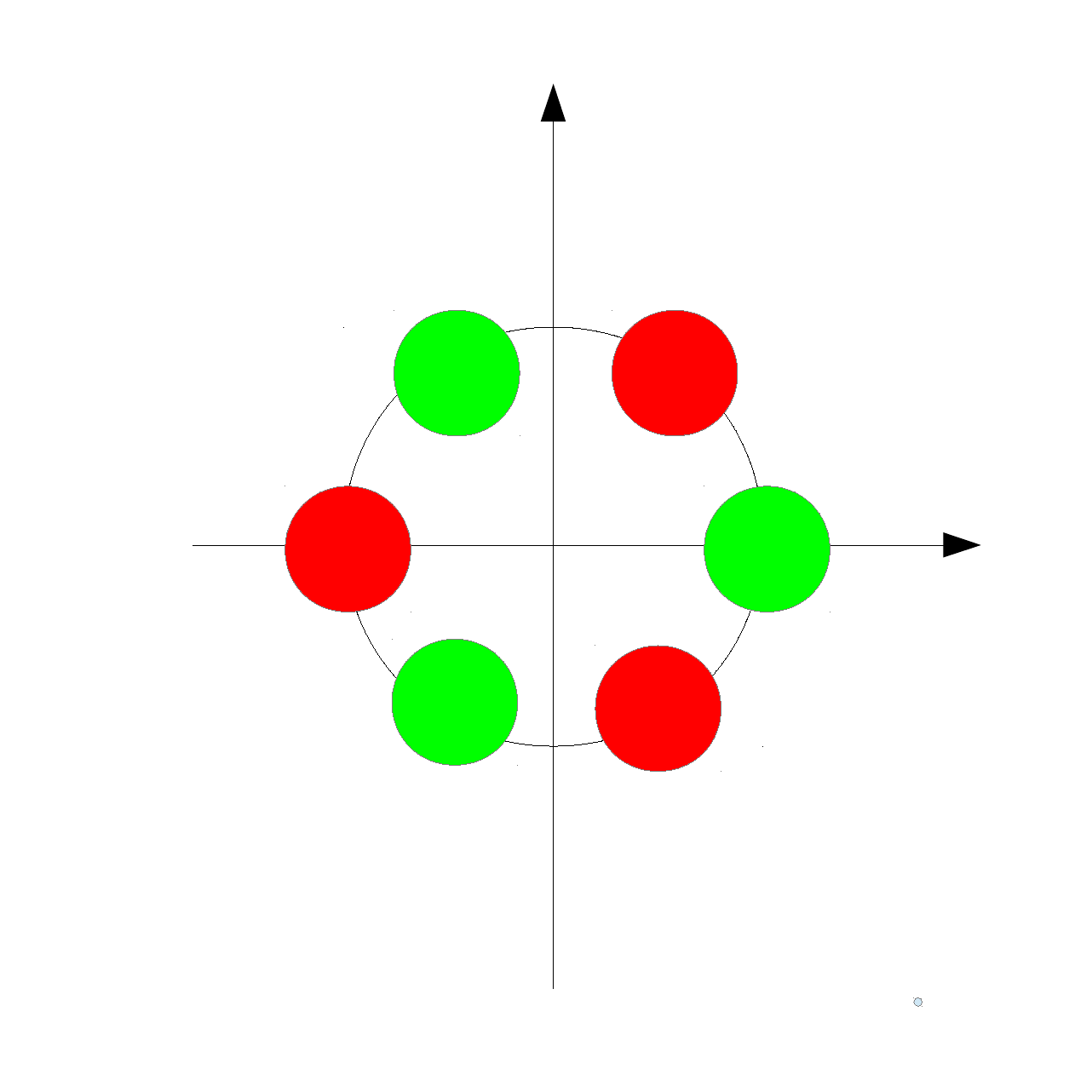}}\quad
  \subfigure[$L=3$]{\includegraphics[width=0.23\textwidth]{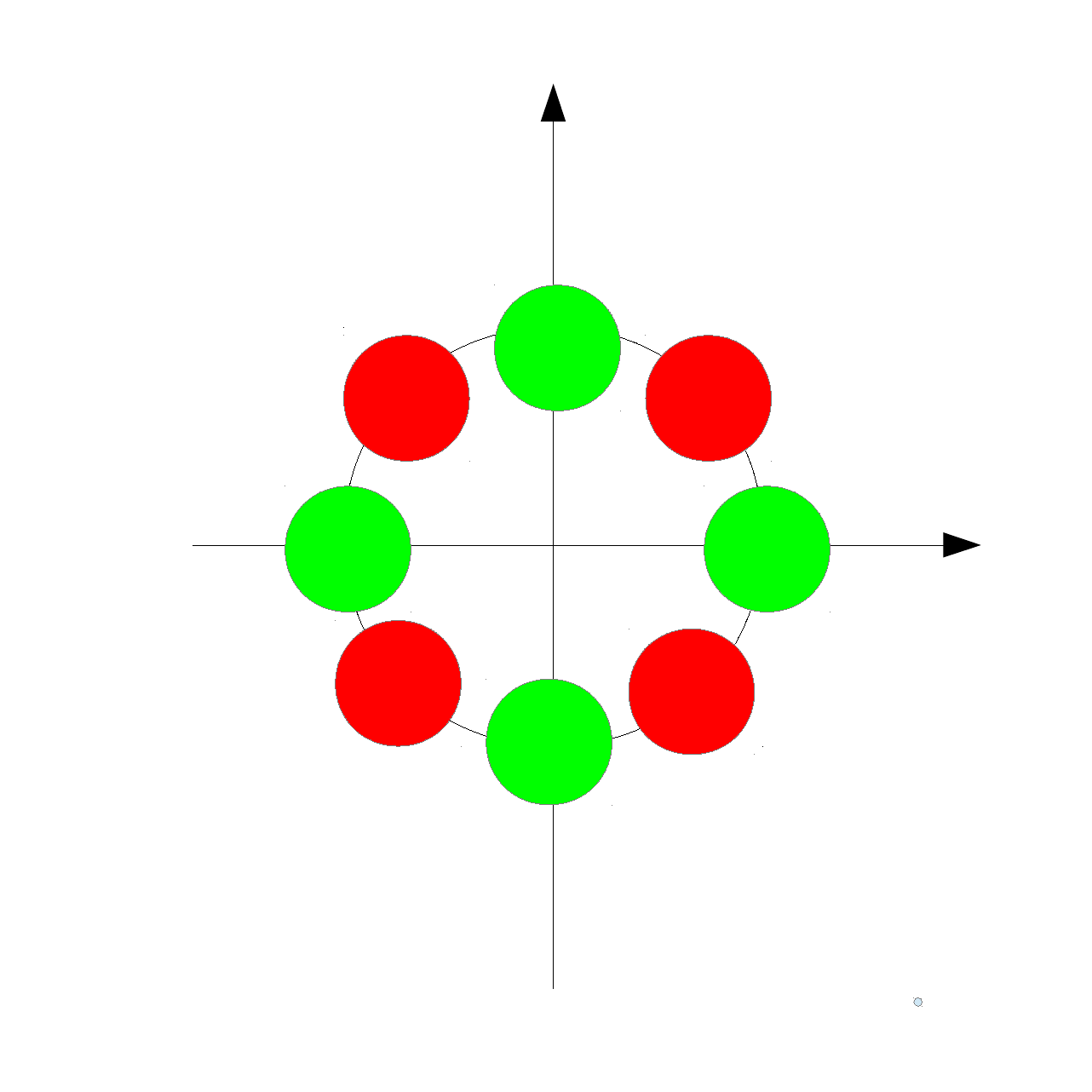}}
  \caption{Illustration of the lowest qubit $(d=2)$ loss codes in phase space. The binary codewords are represented by green
  $(|\bar{0}\rangle)$ and red $(|\bar{1}\rangle)$ circles which are to be superimposed.} 
  \label{fig: phasespace}
\end{figure}

The calculations for the complete, physical AD channel are presented in App. \ref{sec: twoloss}. Besides the basic codewords in the initial code space,
we define the (unnormalised) codewords in all the  three orthogonal subspaces (one code space, and two error spaces for one- and two-photon losses, etc.) as

\begin{equation}
\label{eq: L2notation}
\begin{aligned}
 |\bar{0}_{0}\rangle_{2}&=|\alpha\rangle+|\alpha \exp\left(\frac{2\pi i}{3}\right)\rangle+|\alpha \exp\left(-\frac{2\pi i}{3}\right)\rangle,\\
 |\bar{1}_{0}\rangle_{2}&=|\alpha \exp\left(\frac{\pi i}{3}\right)\rangle+|\alpha \exp\left(\pi i\right)\rangle+|\alpha \exp\left(-\frac{\pi i}{3}\right)\rangle,\\
 |\bar{0}_{1}\rangle_{2}&=|\alpha\rangle+\exp\left(\frac{2\pi i}{3}\right) |\alpha\exp\left(\frac{2\pi i}{3}\right)\rangle
 +\exp\left(-\frac{2\pi i}{3}\right) |\alpha\exp\left(-\frac{2\pi i}{3}\right)\rangle,\\
 |\bar{1}_{1}\rangle_{2}&=|\alpha\exp\left(\frac{\pi i}{3}\right)\rangle+\exp\left(\frac{2\pi i}{3}\right) |\alpha\exp\left(\pi i\right)\rangle
 +\exp\left(-\frac{2\pi i}{3}\right) |\alpha\exp\left(-\frac{\pi i}{3}\right)\rangle,\\
 |\bar{0}_{2}\rangle_{2}&=|\alpha\rangle+\exp\left(-\frac{2\pi i}{3}\right) |\alpha\exp\left(\frac{2\pi i}{3}\right)\rangle
 +\exp\left(\frac{2\pi i}{3}\right) |\alpha\exp\left(-\frac{2\pi i}{3}\right)\rangle,\\
 |\bar{1}_{2}\rangle_{2}&=|\alpha\exp\left(\frac{\pi i}{3}\right)\rangle+\exp\left(-\frac{2\pi i}{3}\right) |\alpha\exp\left(\pi i\right)\rangle
 +\exp\left(\frac{2\pi i}{3}\right) |\alpha\exp\left(-\frac{\pi i}{3}\right)\rangle.
 \end{aligned}
 \end{equation}

 \begin{figure}[t!]
\centering
\includegraphics[width=0.6\textwidth]{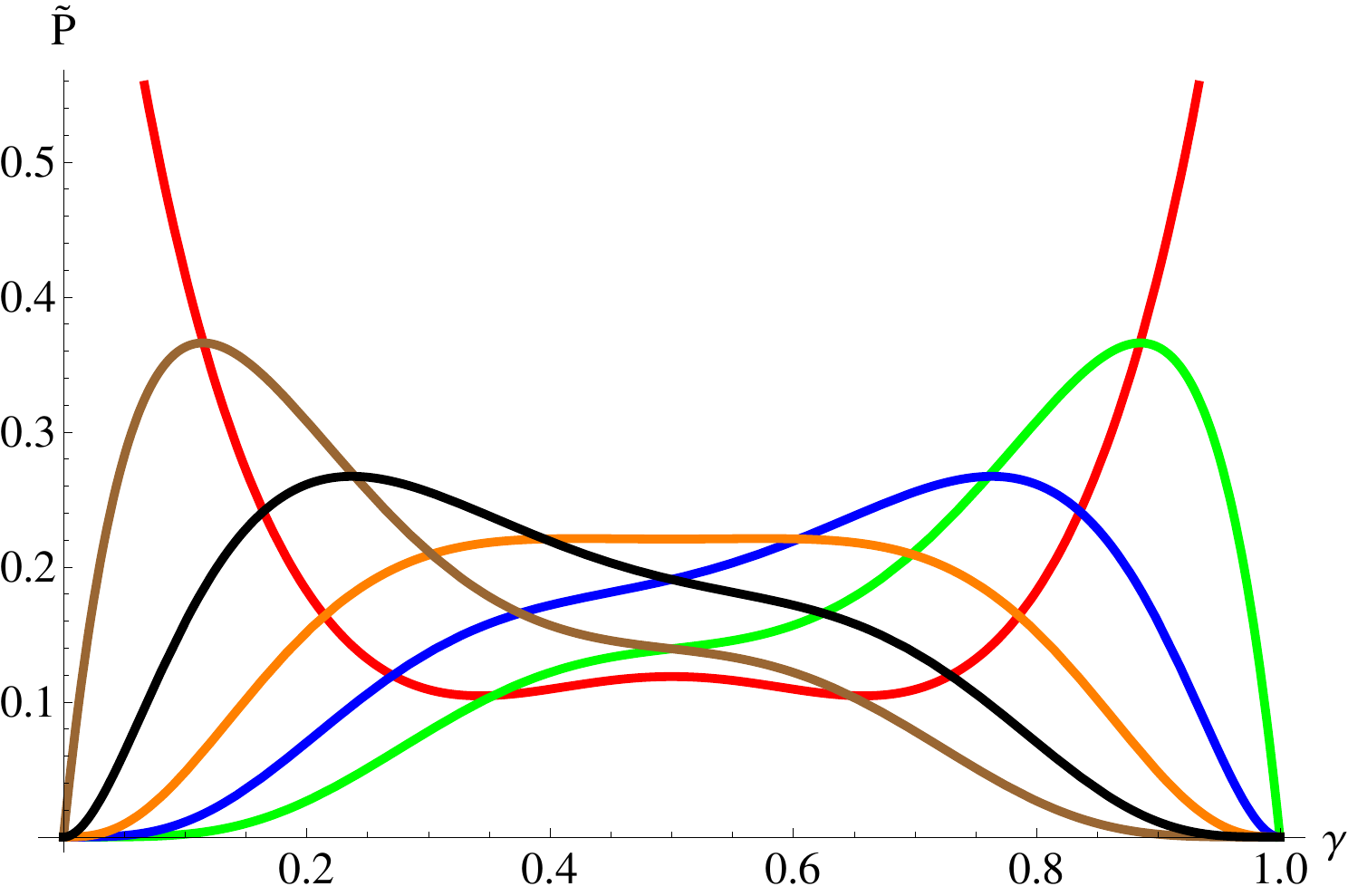}
\caption{Statistical weights with $\alpha=3$ and $a=b=\frac{1}{\sqrt{2}}$ for $L=2$: $\widetilde{p_{0}}$ (red), $\widetilde{p_{1}}$ (green),
$\widetilde{p_{2}}$ (blue),\\ $\widetilde{p_{3}}$ (orange), $\widetilde{p_{4}}$ (black),  $\widetilde{p_{5}}$ (brown) as a function of $\gamma$. Note that the $a,b$-dependence can lead to a different qualitative behaviour
of the probabilities for different logical qubits.}
\label{fig: L2weigths} 
\end{figure}
\noindent
Here, we introduced the notation $\{|\bar{0}_{q}\rangle_{L}, |\bar{1}_{q}\rangle_{L}\}$ to specify the order of the loss code ($L$) and 
the corresponding error space ($q$) ($q=0$ for no loss, $q=1$ for one-photon loss, and $q=2$ for two-photon loss, plus cyclic loss events, see below).
With these definitions for the canonical codewords in the code and error spaces, Eq. \eqref{eq: L2general} simplifies to
\begin{equation}
 \begin{aligned}
\hat{a}^{3k}|\bar{0}_{0}\rangle_{2}&=\alpha^{3k}|\bar{0}_{0}\rangle_{2},\\
\hat{a}^{3k}|\bar{1}_{0}\rangle_{2}&=(-1)^{k}\alpha^{3k}|\bar{1}_{0}\rangle_{2},\\
\hat{a}^{3k+1}|\bar{0}_{0}\rangle_{2}&=\alpha^{3k+1}|\bar{0}_{1}\rangle_{2},\\
\hat{a}^{3k+1}|\bar{1}_{0}\rangle_{2}&=(-1)^{k}\alpha^{3k+1}\exp\left(\frac{\pi i}{3}\right)|\bar{1}_{1}\rangle_{2},\\
\hat{a}^{3k+2}|\bar{0}_{0}\rangle_{2}&=\alpha^{3k+2}|\bar{0}_{2}\rangle_{2},\\
\hat{a}^{3k+2}|\bar{1}_{0}\rangle_{2}&=(-1)^{k}\alpha^{3k+2}\exp\left(\frac{2\pi i}{3}\right)|\bar{1}_{2}\rangle_{2}.
 \end{aligned}
\end{equation}

\begin{figure}[t!]
\centering
\includegraphics[width=0.6\textwidth]{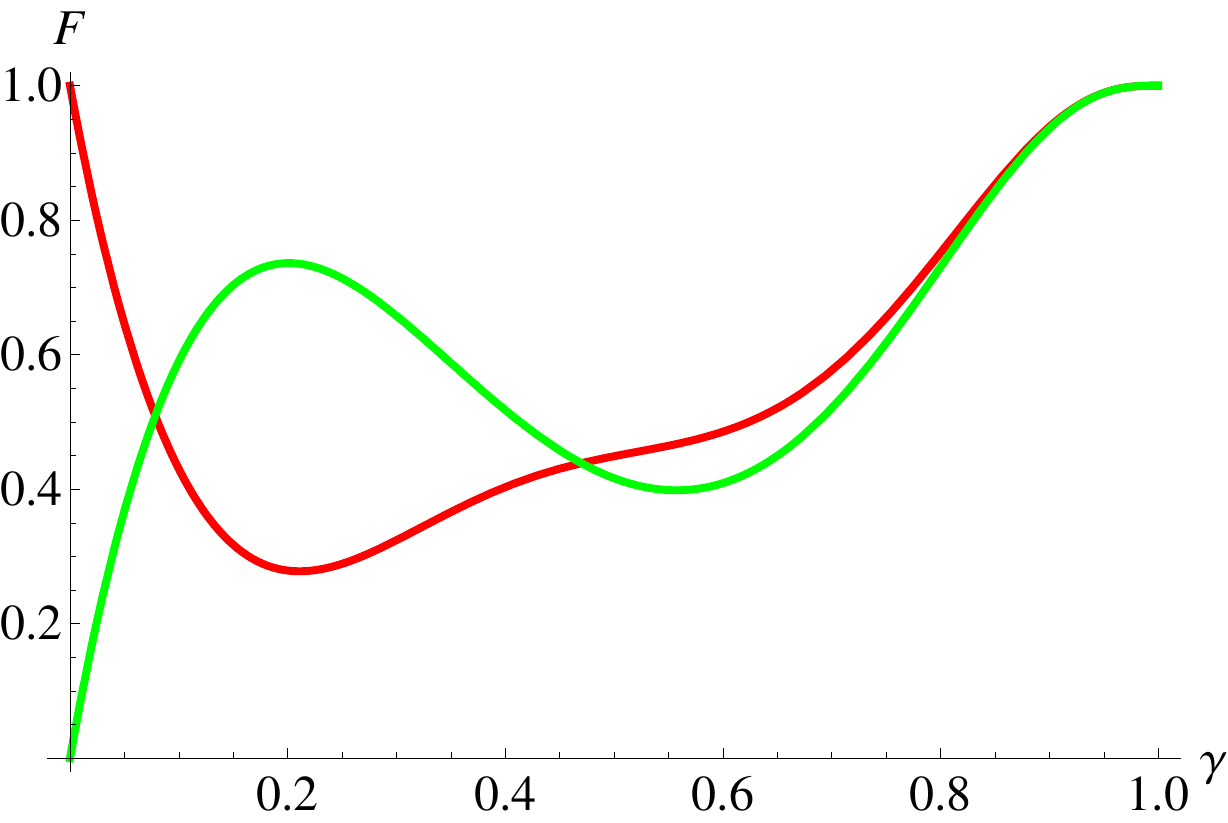}
\caption{Bound $F$ on worst-case fidelity as a function of $\gamma$ for $\alpha=3$ with $a=b=\frac{1}{\sqrt{2}}$ (red) and $a=-b=\frac{1}{\sqrt{2}}$ (green) where $L=2$. The actual lower bound on $F_{wc}$ is given by the minimum of the 
two curves for each $\gamma$.}
\label{fig: L2fid}
\end{figure}

As shown in App. \ref{sec: twoloss}, a logical qubit $a|\bar{0}_{0}\rangle_{2} +b|\bar{1}_{0}\rangle_{2}$ subject to AD becomes a mixture of six components,
which can be cast in the form (omitting proper normalizations of the qubits), 
\begin{equation}
\label{eq: L2mix}
\begin{aligned}
 \bar{\rho} &= p_{0}(a|\widetilde{0}_{0}\rangle_{2}+b|\widetilde{1}_{0}\rangle_{2})\times H.c.\\
 &+p_{1}(a|\widetilde{0}_{1}\rangle_{2}+e^{\frac{i \pi}{3}} b|\widetilde{1}_{1}\rangle_{2})\times H.c.\\
 &+p_{2}(a|\widetilde{0}_{2}\rangle_{2}+e^{\frac{2i \pi}{3}} b|\widetilde{1}_{2}\rangle_{2})\times H.c.\\
 &+p_{3}(a|\widetilde{0}_{0}\rangle_{2}- b|\widetilde{1}_{0}\rangle_{2})\times H.c.\\
 &+p_{4}(a|\widetilde{0}_{1}\rangle_{2}-e^{\frac{i \pi}{3}} b|\widetilde{1}_{1}\rangle_{2})\times H.c.\\
 &+p_{5}(a|\widetilde{0}_{2}\rangle_{2}-e^{\frac{2i \pi}{3}} b|\widetilde{1}_{2}\rangle_{2})\times H.c. ~~.\\
\end{aligned}
\end{equation}
Recall again the additional damping of the amplitude due to the AD channel ($\alpha\rightarrow \sqrt{\gamma}\alpha$) and correspondingly
the adapted notation $\{|\bar{0}_{q}\rangle_{2}, |\bar{1}_{q}\rangle_{2}\}\rightarrow \{|\widetilde{0}_{q}\rangle_{2}, |\widetilde{1}_{q}\rangle_{2}\}$
for $q=0,1,2$. Now the first three terms in Eq. \eqref{eq: L2mix} correspond to correctable logical qubits with, besides some fixed phase gates for $q=1$ and $q=2$,
photon number parities of 0,3,6... or 2,5,8,... or 1,4,7,... corresponding to the loss of 0,6,12,... $(q=0)$ or 1,7,13,...$(q=1)$ or 2,8,14,... $(q=2)$
photons, respectively. The additional terms each mix in uncorrectable phase-flip errors for every subspace corresponding to the loss of 3,9,15,... 
or 4,10,16,... or 5,11,17,... photons. Again, like for the one-loss code, the cyclic behaviour of the simplified model is recovered for the full channel
(for more details, see App. \ref{sec: twoloss}). Thus, for the $L=2$ case, among the dominating loss errors, those from one- and two-photon losses can 
be corrected (i.e., the qubit is still intact in the corresponding error space), whereas those from three-, four- and five-photon losses cannot (i.e., the 
qubit is subject to a phase error). For six- and higher photon losses, the cycle starts again. In general, an $L$-code can correct up to $L$ photon losses
plus other cycles and each codeword has $(L+1)$ coherent-state components living in a $2(L+1)$-dimensional manifold.\\
The lowest cat codes $L=0,1,2,3$ encoding a logical qubit are illustrated in Fig. \ref{fig: phasespace}. In Figs. \ref{fig: L2weigths} and \ref{fig: L2fid}, the statistical weights and the fidelity bound on $F_{wc}$, respectively, are shown as functions of the damping (loss) parameter $\gamma$.

\newpage
\section{Extension To Qudit Codes}
\label{sec: qudit}

\begin{figure}[t!]
  \centering
  \subfigure[$d=8,~ L=0$]{\includegraphics[width=0.3\textwidth]{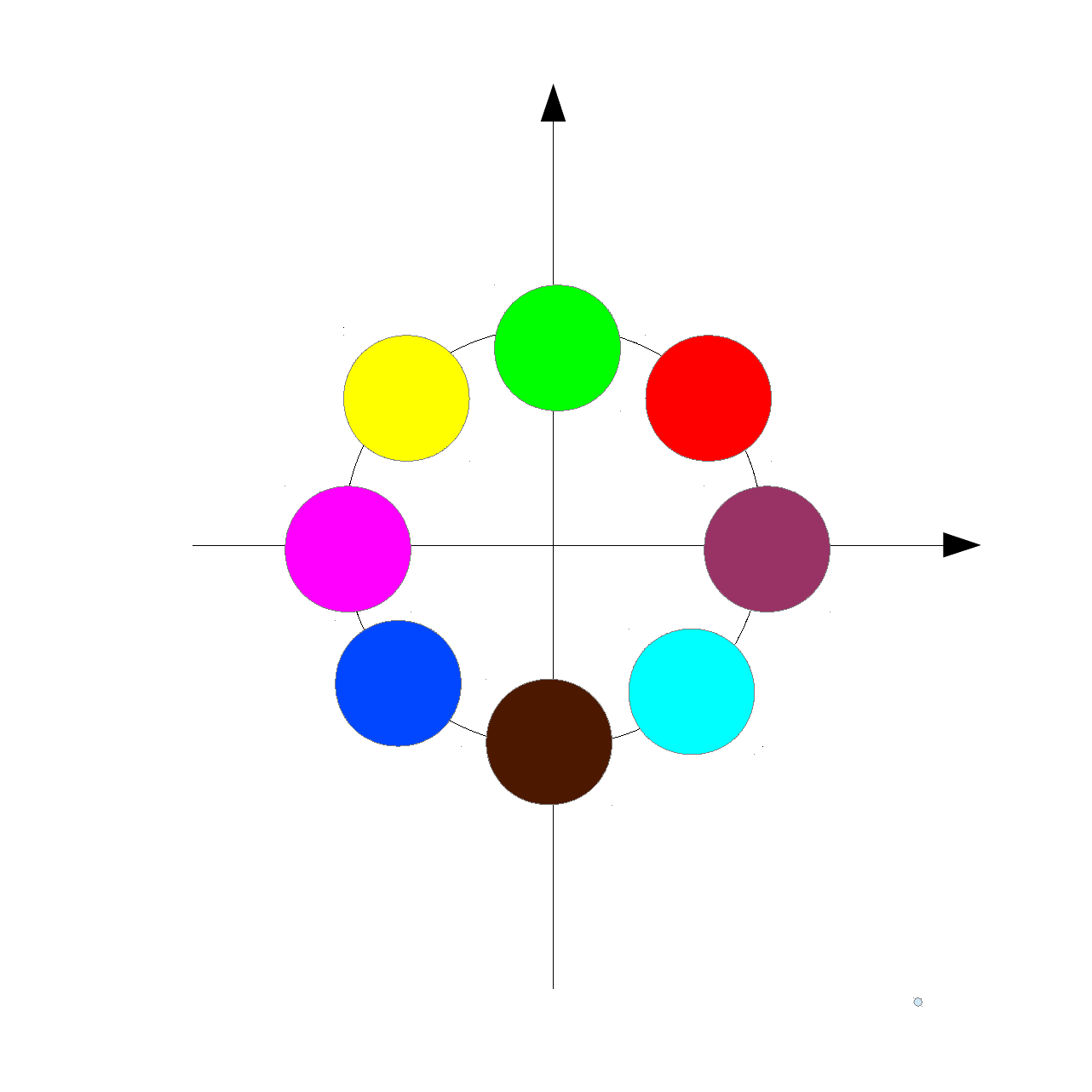}}\quad
  \subfigure[$d=4,~ L=1$]{\includegraphics[width=0.3\textwidth]{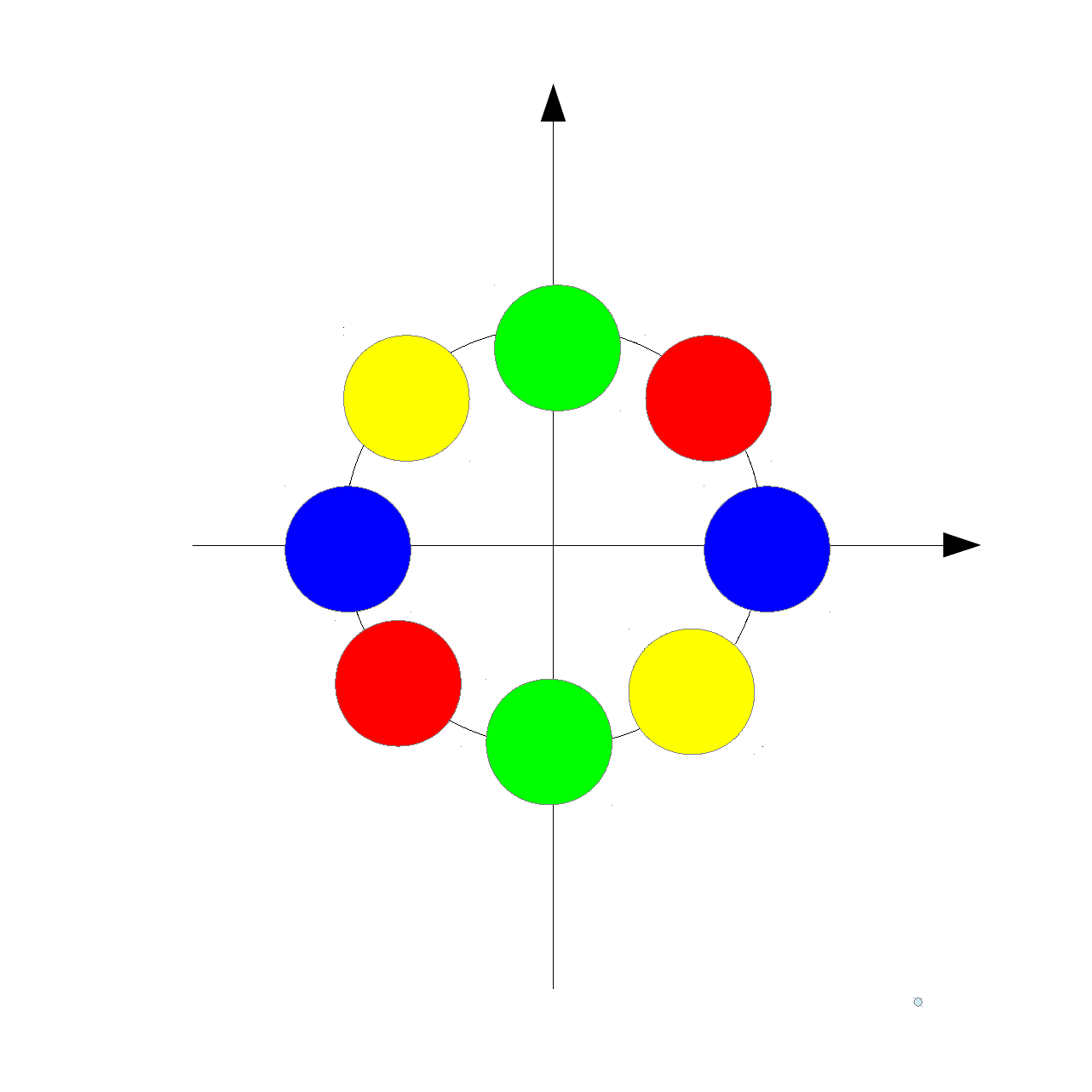}}\quad
  \caption{Illustration of some qudit codes in phase space realized through coherent-state superpositions (every colour indicates 
  another codeword) with in total 8 components (for $d=2,~ L=3$, see Fig 2.)}. 
  \label{fig: qudit-ps}
\end{figure}

Another generalisation that goes beyond the qubit codes presented in the last section is to define equations for the encoding
of an arbitrary qudit of $d$ dimensions:
\begin{equation}
\label{eq: quditdef}
\begin{aligned}
 \exp\left(\frac{2\pi i \hat{n}}{L+1}\right)|\bar{k}\rangle&=|\bar{k}\rangle,\\
(\hat{a}^{L+1}-\exp\left(\frac{2\pi i k}{d}\right)\alpha^{L+1})|\bar{k}\rangle&=0~~\text{for}~~k=0,1,..,d-1 .\\
\end{aligned}
\end{equation}
For $d=2$, Eq. \eqref{eq: definition} for logical qubits is obtained. 
The simplest encoding for general $d$ with $L=0$ corresponds to $|\bar{k}\rangle=|\alpha e^{\frac{2\pi i k}{d}}\rangle$ for $k=0,1,...,d-1$, which is referred to as
'coherent states on a ring' in Ref. \cite{holonomic}.\\
For $d=3$ and $L=1$, i.e. the simplest loss code beyond $d=2$, one finds the (unnormalised) solutions
\begin{equation}
\begin{aligned}
 |\bar{0}\rangle&\equiv |\bar{0}_{+}\rangle=|\alpha\rangle+|-\alpha\rangle,\\
 |\bar{1}\rangle&\equiv |\bar{1}_{+}\rangle=|e^{\frac{i\pi}{3}}\alpha\rangle+|-e^{\frac{i\pi}{3}}\alpha\rangle,\\
 |\bar{2}\rangle&\equiv |\bar{2}_{+}\rangle=|e^{-\frac{i\pi}{3}}\alpha\rangle+|-e^{-\frac{i\pi}{3}}\alpha\rangle.
\end{aligned}
\end{equation}
The three-dimensional code space is spanned by three (generally non-orthogonal) even cat states, similar to the $L=1$ qubit code which has two even cat states as codewords.
In the simplified error model, we also find a similar cyclic behaviour of the codewords, 

\begin{equation}
\begin{aligned}
\hat{a}^{2k} |\bar{0}\rangle&=\alpha^{2k}|\bar{0}_{+}\rangle,\\
\hat{a}^{2k}|\bar{1}\rangle&=\exp\left(\frac{i\pi}{3}\right)^{2k} \alpha^{2k}|\bar{1}_{+}\rangle,\\
\hat{a}^{2k}|\bar{2}\rangle&=\exp\left(-\frac{i\pi}{3}\right)^{2k}\alpha^{2k}|\bar{2}_{+}\rangle,\\
\hat{a}^{2k+1} |\bar{0}\rangle&=\alpha^{2k+1}|\bar{0}_{-}\rangle,\\
\hat{a}^{2k+1}|\bar{1}\rangle&=\exp\left(\frac{i\pi}{3}\right)^{2k+1}\alpha^{2k+1}|\bar{1}_{-}\rangle,\\
\hat{a}^{2k+1}|\bar{2}\rangle&=\exp\left(-\frac{i\pi}{3}\right)^{2k+1}\alpha^{2k+1}|\bar{2}_{-}\rangle,\\ 
\end{aligned}
\end{equation}
where $|\bar{0}_{-}\rangle=|\alpha\rangle-|-\alpha\rangle, |\bar{1}_{-}\rangle=|e^{\frac{i\pi}{3}}\alpha\rangle-|-e^{\frac{i\pi}{3}}\alpha\rangle$
and $|\bar{2}_{-}\rangle=|e^{-\frac{i\pi}{3}}\alpha\rangle-|-e^{-\frac{i\pi}{3}}\alpha\rangle$.
Depending on the number of lost photons $m=0,1,2,3,4,5$, the logical qudit $a|\bar{0}\rangle +b|\bar{1}\rangle+c|\bar{2}\rangle$ suffers from random relative phases
(the phase factors $(e^{\pm \frac{i\pi}{3}})^{2k}$ in front of the transformed codewords $|\bar{1}_{\pm}\rangle$ and $|\bar{2}_{\pm}\rangle$). In fact, only for 
$k=0,3,6,...$ no phase errors occur and a fixed phase gate (the phase factor $e^{\pm \frac{i\pi}{3}}$ in front of  $|\bar{1}_{-}\rangle$ and $|\bar{2}_{-}\rangle$)
is either applied (1,7,13,.. losses) or not (0,6,12,...losses). Subject to the full AD channel, the mixed output state for a logical qutrit $a|\bar{0}\rangle +b|\bar{1}\rangle+c|\bar{2}\rangle$ has six components,
\begin{equation}
\begin{aligned}
|\bar{\psi}_{0}\rangle&=a|\bar{0}_{+}\rangle+b|\bar{1}_{+}\rangle+c|\bar{2}_{+}\rangle,\\
|\bar{\psi}_{1}\rangle&=a|\bar{0}_{-}\rangle+b\exp\left(\frac{i\pi}{3}\right)|\bar{1}_{-}\rangle+c\exp\left(-\frac{i\pi}{3}\right)|\bar{2}_{-}\rangle,\\
|\bar{\psi}_{2}\rangle&=a|\bar{0}_{+}\rangle+b\exp\left(\frac{2i\pi}{3}\right)|\bar{1}_{+}\rangle+c\exp\left(-\frac{2i\pi}{3}\right)|\bar{2}_{+}\rangle,\\
|\bar{\psi}_{3}\rangle&=a|\bar{0}_{-}\rangle-b|\bar{1}_{-}\rangle-c|\bar{2}_{-}\rangle,\\
|\bar{\psi}_{4}\rangle&=a|\bar{0}_{+}\rangle-b\exp\left(\frac{i\pi}{3}\right)|\bar{1}_{+}\rangle-c\exp\left(-\frac{i\pi}{3}\right)|\bar{2}_{+}\rangle,\\
|\bar{\psi}_{5}\rangle&=a|\bar{0}_{-}\rangle-b\exp\left(\frac{2i\pi}{3}\right)|\bar{1}_{-}\rangle-c\exp\left(-\frac{2i\pi}{3}\right)|\bar{2}_{-}\rangle,
\end{aligned}
\end{equation}
with some statistical weights. Here, only $|\bar{\psi}_{0}\rangle$ and $|\bar{\psi}_{1}\rangle$ correspond to correctable qutrits (corresponding
to 0,6,12,... and 1,7,13,.. losses, respectively). All the remaining qutrits $|\bar{\psi}_{2}\rangle,|\bar{\psi}_{3}\rangle, |\bar{\psi}_{4}\rangle$
and $|\bar{\psi}_{5}\rangle$ have suffered from phase errors (corresponding to 2,8,14,... or 3,9,15... or 4,10,16,... or 5,11,17,.. losses, respectively). After six
losses a new cycle starts.\\  
In general, for a general $L$-code the period of a cycle depends on the total number of coherent-state components of the code (that is $d(L+1)$), e.g. a 4-cycle (i.e., 4 terms in $\bar{\rho}$)
for $d=2|L=1$ or a 6-cycle (6 terms in $\bar{\rho}$) for both $d=2|L=2$ and $d=3|L=1$.
\section{Application In A One-Way Quantum Communication Scheme}
\label{sec: oneway communication}
\begin{figure}[h!]
  \centering
  \subfigure["old" scheme]{\includegraphics[width=0.6\textwidth]{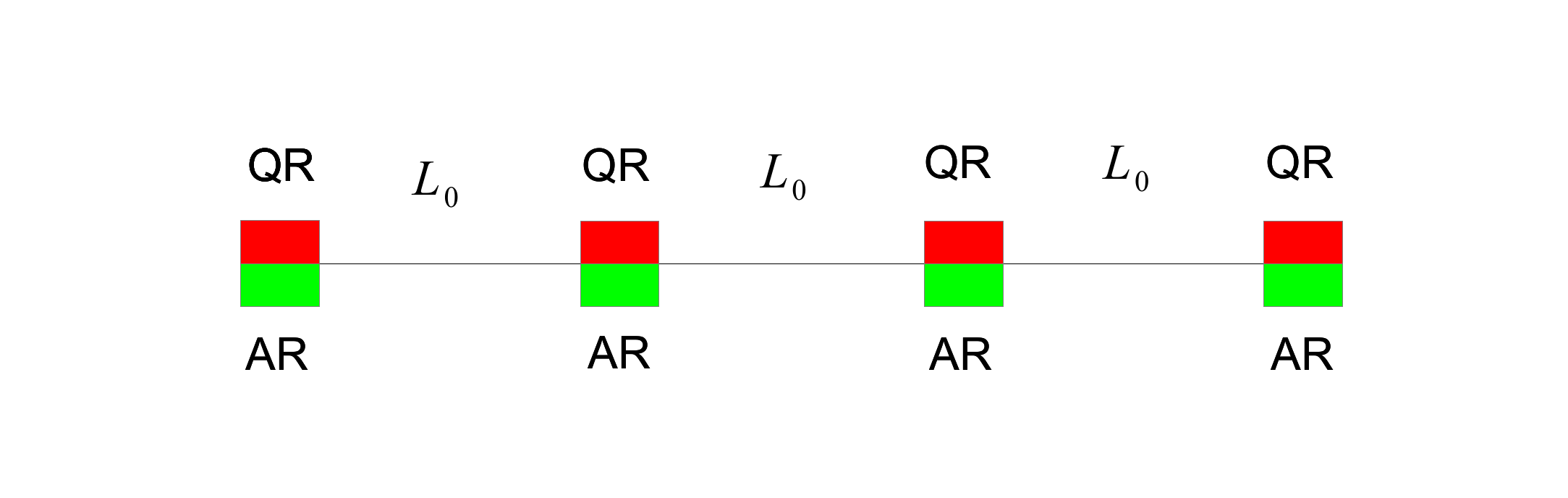}}\quad
  \subfigure[improved, "new" scheme]{\includegraphics[width=0.6\textwidth]{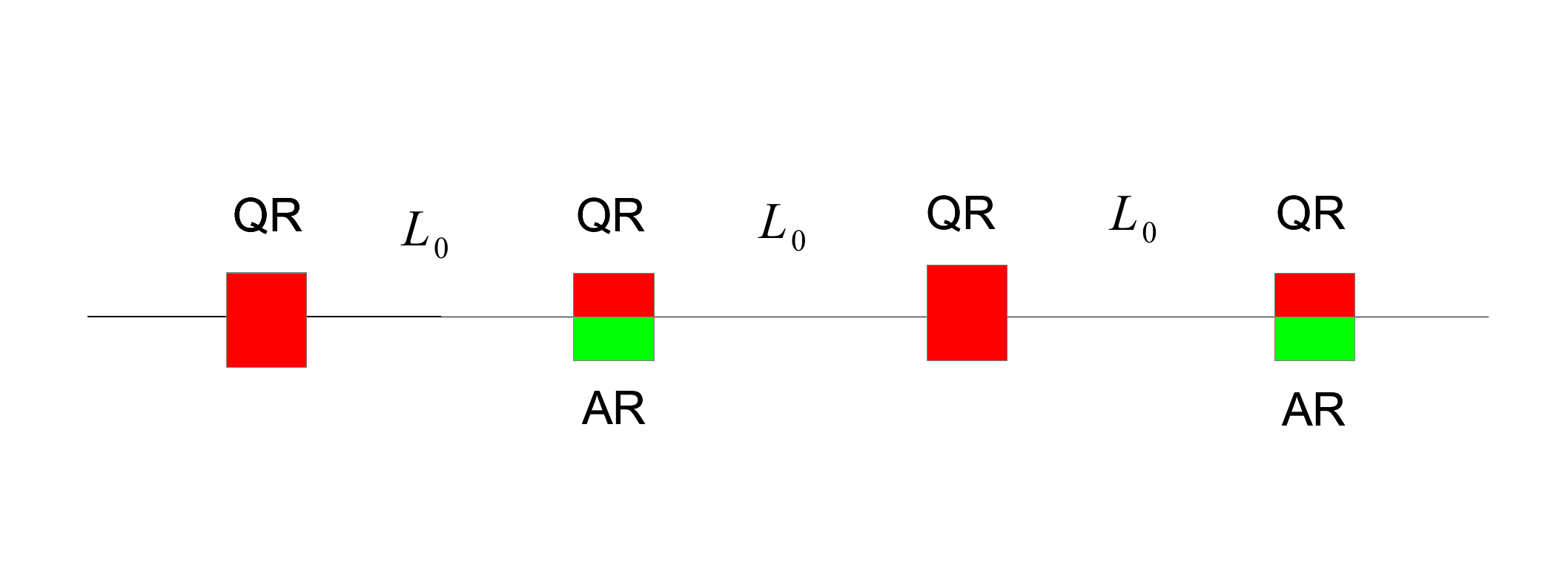}}\quad
  \caption{Schematic of a one-way quantum repeater with qubit recovery (QR, red) as well as amplitude restoration (AR, green) at every
  repeater station (a) or with QR at every repeater station and AR only at every second station (b).} 
  \label{fig: newold}
\end{figure}
\newpage
Loss-adapted quantum error correction codes are a key ingredient for a so-called third-generation quantum repeater \cite{QPC}. In such a third-generation quantum repeater, 
the goal is to transmit an encoded qubit over a total distance  $\mathcal{L}$ without distributing, as an initial step, entangled states over smaller segments
of the entire channel like in a more standard quantum repeater (i.e. either a first-generation repeater based on entanglement purification and swapping \cite{Briegel, Duer}
or a second-generation one that includes quantum error correction of local errors \cite{JiangQR}).
\begin{figure}[t!]
  \centering
  \subfigure[short scale]{\includegraphics[width=0.45\textwidth]{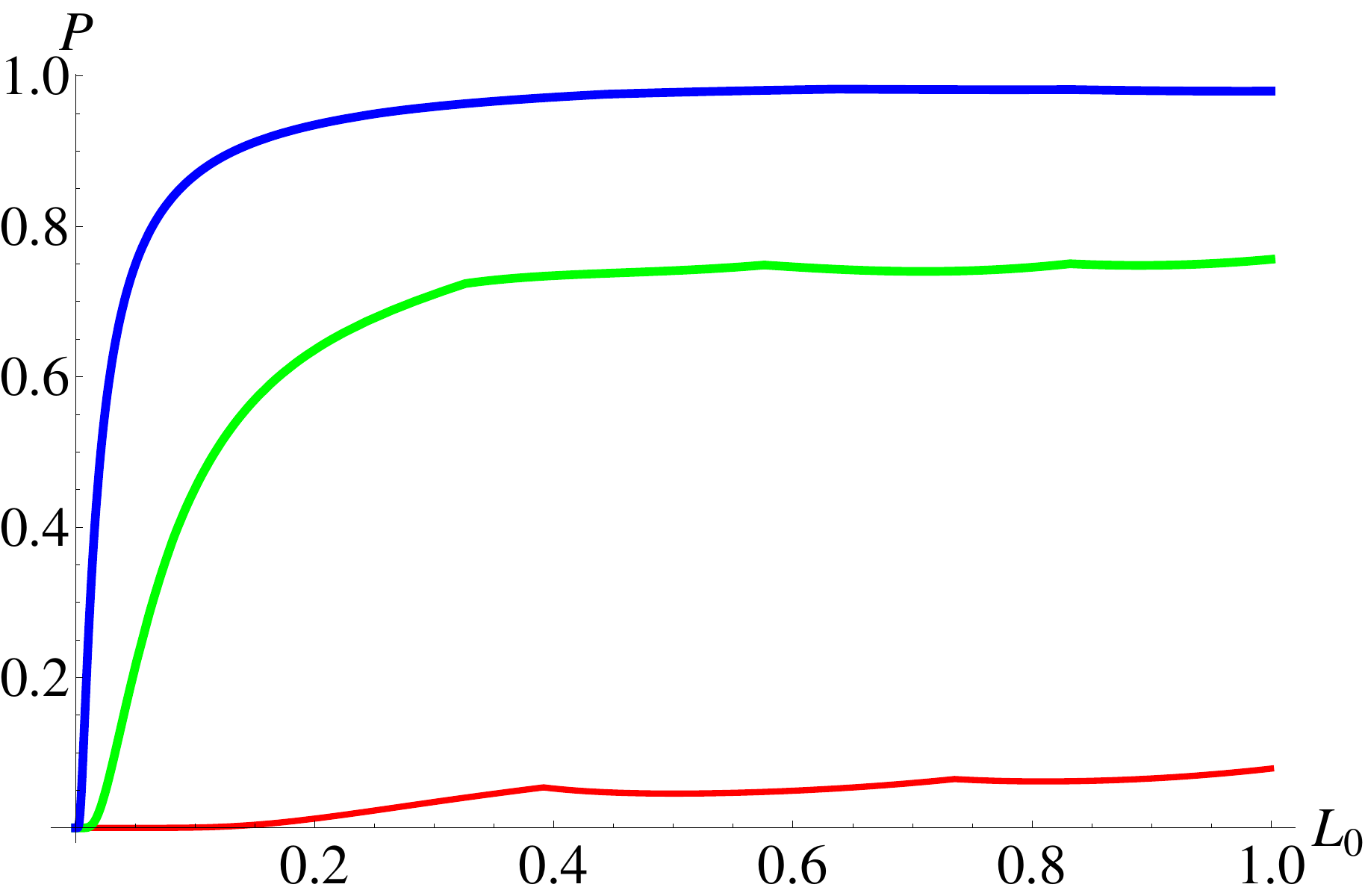}}\quad
  \subfigure[long scale]{\includegraphics[width=0.45\textwidth]{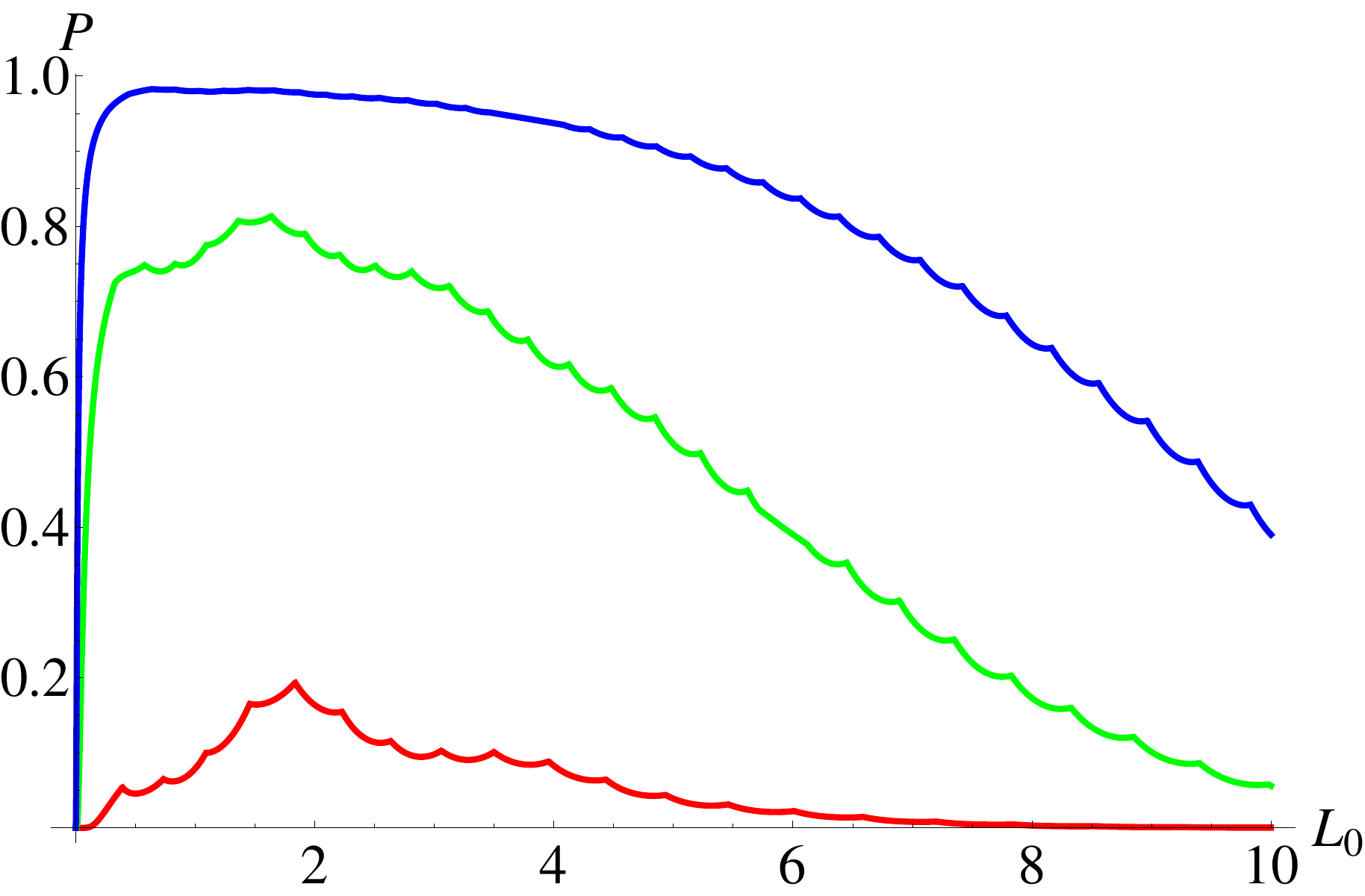}}\quad
  \caption{Total success probability $P$ of amplitude restoration as a function of the elementary distance $L_{0}$ in an improved ("new") one-way scheme over
a total distance of 1000 km for $L=4$ and various $\alpha$: $\alpha=6$ (red), $\alpha=7$ (green), $\alpha=8$ (blue).} 
  \label{fig: longshort}
\end{figure}
Nonetheless, also in a third-generation repeater, the total distance $\mathcal{L}$ is divided into smaller elementary distances
$L_{0}<\mathcal{L}$ and at each distance $L_{0}$ a repeater station is placed. However, at every repeater station, quantum error correction
(especially in order to suppress the photon transmission loss) is performed on an incoming 
qubit which has travelled over the distance $L_{0}$. The recovered logical qubit is then sent further to the next station and so on until
it reaches the final repeater station. Thus, in a third-generation repeater, there is no need to temporarily store entangled states until neighbouring 
entangled states have been distributed and purified, and there is also no need to send classical information back and forth between repeater stations. 
Such two-way classical communication slows down the repeater (and hence reduces the rate) and it also makes good quantum memories a necessity. 
In contrast, a third-generation repeater only requires one-way classical communication and, in principle, no quantum memories are needed at all. 
Quantum information is sent directly at rates that approach, in principle, those achievable in classical communication. As loss-protected qubits are usually
encoded into multi-mode states \cite{QPC}, an attractive feature of the cat loss code would be that only a single optical mode must be sent.\\ 
In the case of cat codes, the first step at each repeater station is a QND-type parity measurement that  determines the 
corresponding error space. After fixing the parity, the logical state is recovered to a great extent and the initial logical qubit resides in some error space with high, but non-unit fidelity.\\
As mentioned in the former sections, a special problem that occurs with the transmission of cat-code qubits is the distance-dependent damping of the 
amplitude. In addition to the qubit recovery (QR) at each repeater station, the amplitude has to be restored as well. A probabilistic scheme for this amplitude restoration (AR) is presented in App. \ref{sec: restauration}. 
In our AR scheme, we use quantum teleportation and choose to teleport the qubit back into the code space, while restoring the amplitude.    
A schematic is depicted in Fig. \ref{fig: newold} a). After each repeater station, the qubit is recovered as well as the amplitude is restored.
Figure \ref{fig: newold} b) shows an improved scheme in which the qubit is still recovered at each repeater station, but the amplitude is restored at 
every second repeater station only. The total success probability for this improved scheme is shown in Fig. \ref{fig: longshort}.
One observes that the total success probability initially increases with the elementary distance $L_{0}$ before reaching a maximum and tending
to zero again. Indeed, doing AR at the end of the total channel at distance $\mathcal{L}$ corresponds to an exponentially small success probability, while
a scheme in which AR is performed too frequently also means that the probabilistic element introduced via AR accumulates over the total distance. We expect that 
a further improvement compared to the results shown in Fig. \ref{fig: longshort} can be obtained by doing AR even less frequently than 
at every second repeater station. Here, we shall only demonstrate an in-principle improvement when QR and AR are not always performed synchronously, 
without intending to find an optimal scheme. The fidelity, however, is near unity for short elementary distances and decreases with increasing $L_{0}$ (see Fig. \ref{fig: owfidelity}).\\ 
To summarise, qubit recovery is necessary after sufficiently short distances, whereas amplitude restoration seems to be beneficial after longer but not too long distances.
That the logical qubits must be recovered frequently after short distances is also expected, since the loss code does not tolerate too large losses
for the quantum information to remain intact. A comparison of the success probabilities and fidelities for the "old" and the improved "new" scheme with different cat codes and different amplitudes 
is shown in Tables \ref{table: L=3}-\ref{table: L=5}. Besides the significantly higher success probabilities, the improved scheme also gives slightly better fidelities.\\
In general, the expected trade-off is recovered: for too large amplitudes $\alpha$, the photon loss probability goes up (and hence the fidelity decreases) 
while the codewords become more orthogonal (and hence the filter, see App. \ref{sec: restauration}, and thus the AR probabilities increase). Conversely, for smaller $\alpha$,
the AR becomes less likely to succeed, while larger fidelities can be obtained. A non-trivial result is to find a code $L$ and a protocol, for which
an $\alpha$-regime exists that allows for both reasonable success probabilities ($\sim 1\%-10\%$) and near-unit fidelities at some elementary distances $L_{0}$.
For $L=3$ using the "old" scheme such an $\alpha$-regime does not seem to exist (see Table \ref{table: L=3}). With the "new", improved scheme, however, the $L=3$-code may suffice
for elementary distances of $L_{0}\sim 100~\text{m}$. For the $L=4$- and $L=5$-codes, both schemes can work at elementary distances
of $L_{0}\sim 10- 100~\text{m}$. A general observation is that elementary distances as large as $\sim 1~\text{km}$ result in very bad fidelities. Thus, a cat-encoded
logical qubit is more sensitive to too large losses and too large $L_{0}$ than, for instance, a single-photon-based, multi-mode, QPC-encoded qubit for which
$L_{0}\sim 1~\text{km}$ works \cite{QPC, Fab}. However, the fact that a cat-encoded qubit only requires a single optical mode means that low success probabilities
in a single repeater chain could be efficiently compensated via (e.g. broadband) parallelisation or multiplexing.

\begin{figure}[t!]
\centering
\includegraphics[width=0.6\textwidth]{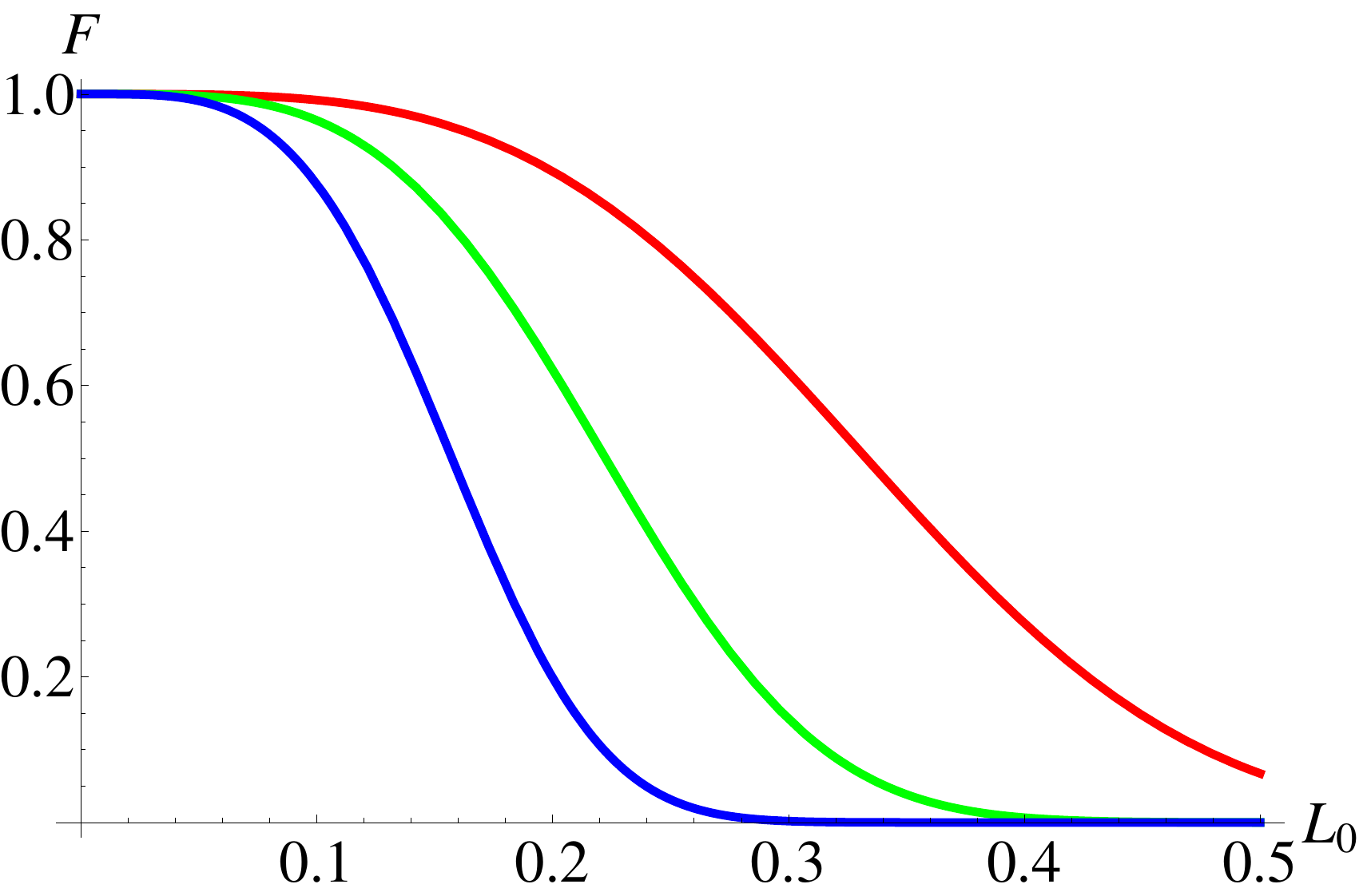}
\caption{Bound $F$ on worst-case fidelity as a function of the elementary distance $L_{0}$ for a one-way scheme over
a total distance of 1000 km with $L=4$ for various $\alpha$: $\alpha$: $\alpha=6$ (red) $\alpha=7$ (green) $\alpha=8$ (blue).}
\label{fig: owfidelity}
\end{figure}

\begin{table}[b!]
\begin{tabular}{|c|c|c|c|c|c|}
\hline
$\alpha$ &	$L_{0}$	 &       $F_{new}$ &	   $P_{new}$	&  $F_{old}$&  $P_{old}$\\
\hline
4.0&	0.01&	0.99998900&	$\approx 0$&	0.999989&	$\approx 0$\\
\hline
4.0&	0.10&	0.98944600&	$\approx 0$&	0.989275&	$10^{-76}$\\
\hline
4.0&	1.00&	0.00473919&	$3\cdot 10^{-8}$&	0.00232537&	$10^{-12}$\\
\hline
 4.5&	0.01&	0.99997000&	$\approx 0$&	0.99997&	$10^{-42}$\\
\hline
\rowcolor{yellow}4.5&	0.10&	0.97327800&	0.00830884&	0.972789&	$7\cdot 10^{-5}$\\
\hline
4.5&	1.00&	$9\cdot 10^{-6}$&	0.00880618&	$10^{-6}$&	$3\cdot 10^{-3}$\\
\hline
5.0&	0.01&	0.99993100&	$\approx 0$&	0.999931&	$10^{-67}$\\
\hline
5.0&	0.10&	0.94012200&	$5\cdot 10^{-4}$&	0.87604&	$2\cdot 10^{-7}$\\
\hline
5.0&	1.00&	$\approx 0$&	0.16894200&	$6\cdot10^{-22}$&	0.0847453\\
\hline
6.0&	0.01&	0.99970600&	$5\cdot 10^{-4}$&	0.999705&	$3\cdot 10^{-7}$\\
\hline
6.0&	0.10&	0.77462700&	0.46871500&	0.771153&	0.221926\\
\hline
6.0&	1.00&	$\approx 0$&	0.89348900&	$3\cdot 10^{-36}$&	0.843821\\
\hline
\end{tabular}
\caption{Comparison between the "old" and the "new" schemes for a total distance of $\mathcal{L}=1000~\text{km}$ with the $L=3$-code, $L_{0}$ in km. Colour indicates
near-feasible regimes.}
\label{table: L=3}
\end{table}

\begin{table}[t!]
\begin{tabular}{|c|c|c|c|c|c|}
\hline
$\alpha$&	$L_{0}$	&        $F_{new}$ &	   $P_{new}$	&  $F_{old}$ & $P_{old}$\\
\hline
6&	0.01&	0.999999&	$3\cdot10^{-31}$&	0.999999&	$10^{-61}$\\
\hline
6&	0.1&	0.991757&	$4\cdot 10^{-4}$&	0.991574&	$4\cdot 10^{-7}$\\
\hline
6&	1.00&	$10^{-9}$&	0.0787418&	$4\cdot 10^{-11}$&	0.0455329\\
\hline
7&	0.01&	0.999996&	$6\cdot 10^{-4}$&	0.999996&	$3\cdot 10^{-7}$\\
\hline
\rowcolor{yellow}7&	0.1&	0.963915&	0.451687&	0.96314&	0.214877\\
\hline
7&	1.00&	$6\cdot 10^{-28}$&	0,755955&	$3\cdot 10^{-22}$&	0.740854\\
\hline
\rowcolor{yellow}8&	0.01&	0.999983&	0.230988&	0.999983&	0.0531004\\
\hline
8&	0.1&	0.876309&	0.867937&	0.873809&	0.74901\\
\hline
8&	1.00&	$10^{-66}$&	        0.979637&	$10^{-75}$&	        0.977982\\
\hline
\end{tabular}
\caption{Comparison between the "old" and the "new" schemes for a total distance of $\mathcal{L}=1000~\text{km}$ with the $L=4$-code, $L_{0}$ in km.
Colour indicates feasible regimes.}
\label{table: L=4}
\end{table}

\begin{table}[t!]
\begin{tabular}{|c|c|c|c|c|c|}
\hline
$\alpha$&	$L_{0}$	 &       $F_{new}$ &	   $P_{new}$	&  $F_{old}$ & $P_{old}$\\
\hline
6&	0.01&	 $\approx 1$&	             $\approx 0$&	              $\approx 1$&	    $\approx 0$\\
\hline
6&	0.1&	0.999781&	$4\cdot 10^{-24}$&	        0.999776&	$10^{-47}$\\
\hline
6&	1.00&	0.00639287&	$3\cdot10^{-5}$&	$3\cdot 10^{-3}$&	$6\cdot 10^{-6}$\\
\hline
7&	0.01&	 $\approx 1$&	         $3\cdot 10^{-50}$&	     $\approx 1$&	        $\approx0$\\
\hline
7&	0.1&	0.998659&	$\cdot10^{-5}$ &	0.998624&	$10^{-10}$\\
\hline
7&	1.00&	$2\cdot 10^{-9}$&	0.06615&	$4\cdot 10^{-11}$ & 0.0759747\\
\hline
8&	0.01&	 $\approx 1$&	        $10^{-7}$&	 $\approx 1$&	        $2\cdot 10^{-14}$\\
\hline
\rowcolor{yellow}8&	0,1&	0.99371&	0.194448&	0.993546&	0.0417406\\
\hline
8&	1.00&	$4\cdot 10^{-27}$&	0.691036&	$10^{-31}$&	0.659869\\
\hline
9&	0.01&	$\approx 1$&	        $4\cdot10^{-3}$&	$\approx 1$& $1.75\cdot 10^{-5}$\\
\hline
\rowcolor{yellow}9&	0.1&	0.975983&	0.578119&	0.97537& 0.334447\\
\hline
9&	1.00&	$5\cdot 10^{-64}$&	0.963224&	$4\cdot 10^{-74}$& 0.89103\\
\hline
\end{tabular}
\caption{Comparison between the "old" and the "new" schemes for a total distance of $\mathcal{L}=1000~\text{km}$ with the $L=5$-code, $L_{0}$ in km.
Colour indicates feasible regimes.}
\label{table: L=5}
\end{table}

\section{Conclusions}
We analysed a generalised quantum error correction code that is adapted to correct errors induced from photon losses and is based on superpositions of coherent states.
Our generalisation includes instances of such a cat code where errors from more than one photon loss can be, in principle, approximately corrected. 
For the higher loss codes, however, the overlap of the codewords increases and must be compensated by an increasing coherent-state amplitude which results
in a growing error rate. Thus, one encounters the usual trade-offs when a continuous-variable encoding is employed. We illustrate such an effect for the example
of a one-way quantum communication scheme for large distances based on cat codes.\\
The non-orthogonality of the codewords could be entirely avoided by choosing a particular logical basis in the code space (the $\bar{X}$- instead
of the $\bar{Z}$-basis), however, this would be at the expense of a deformation of the logical qubits for finite coherent-state amplitudes leading
to a complicated and undesirable output density matrix. Our choice of the $\bar{Z}$-basis circumvents this deformation at the expense of non-zero codeword
overlap.\\
Another generalisation that we discussed for cat codes is for a higher-dimensional code space beyond logical qubits, i.e., qudits. Future work will aim at
potential optical implementations of these codes, including practical ways to encode, to do the measurements (parity detections), and to achieve the coherent-state 
amplitude restorations (either by creating the encoded, entangled ancilla states, as proposed here, or by employing an alternative method).
\section{Acknowledgement}
\noindent We acknowledge support from Q.com (BMBF) and Hipercom (ERA-NET CHISTERA).

\bibliographystyle{apsrev4-1}
\bibliography{Catcode.bib}

\newpage
\appendix

\section{Knill-Laflamme conditions For QEC}
\label{sec: QEC}
Decoherence is a usually undesired quantum effect that prevents a quantum system from a purely unitary evolution.
One method to overcome decoherence is the usage of a quantum error correction code. A quantum error correction code is a 
$d$-dimensional subspace of some higher-dimensional Hilbert space. The $d$ basis vectors $\{|c_{1}\rangle,|c_{2}\rangle,...,|c_{d}\rangle\}$ are referred to as
codewords and any normalised 
superposition corresponds to a logical or an encoded qudit.\\
Given an explicit error model with error operators $E_{i}$, it can be shown that the action of certain error operators
from the so-called correctable set can be reversed by means of a recovery operation, if and only if the following two sets of conditions are fulfilled:
The first set of condition states that corrupted codewords are orthogonal
\begin{equation}\label{ortho}
 \langle c_{k}|E_{i}^{\dagger}E_{j}|c_{l}\rangle =0~~ \text{if}~~ k\neq l,
\end{equation}
which partially incorporates the quantum mechanical requirement for distinguishability of the different code and error spaces.
The second one includes the non-deformability condition, i.e.
\begin{equation}\label{deform}
  \langle c_{l}|E_{i}^{\dagger}E_{i}|c_{l}\rangle =g_{i},~~ \forall l,
\end{equation}
which states that the norm of corrupted state only depends on the error and not on the codeword. These two sets of conditions are referred 
to as the Knill-Laflamme conditions. An encoding that exactly fulfils the KL conditions for a set of errors is called an exact code and the 
corresponding set is the correctable error set.\\
In optical quantum information, the main mechanism of decoherence is photon loss, as described in Section \ref{sec: cat-code}. In this context, 
the notion of approximate quantum error correction codes has been introduced \cite{approximate_codes}. In an approximate QEC for AD, 
the KL conditions are only fulfilled up to a certain order in the loss parameter $1-\gamma$, thus orthogonality and non-deformability
are not strictly fulfilled. In the context of the present cat codes, it also depends on the amplitude $\alpha$ whether the KL conditions are fulfilled
or not.

\section{Full loss channel and KL conditions for the one-loss cat code}
\label{sec: oneloss}

Let us first consider the simplified set of errors $\mathcal{E}=\{\hat{a}^{4k},\hat{a}^{4k+1}, k\in \mathbb{N}_{0}\}$. 
For the even-cat codewords of Eq. \eqref{eqn: catcode} (i.e., the ``$\bar{Z}$-basis''), we have the following KL conditions:
\begin{equation}
\begin{aligned}
\langle\bar{0}_{+}|(\hat{a}^{4k})^{\dagger}\hat{a}^{4k}|\bar{1}_{+}\rangle= &i^{4k}(\alpha^2)^{4k}\frac{1}{\sqrt{N_{+}}}(\langle \alpha|+\langle -\alpha|)\cdot\frac{1}{\sqrt{N_{+}}}(|i\alpha\rangle+|-i\alpha\rangle)\\
 &=\frac{(\alpha^2)^{4k}}{N_{+}}\left(\langle \alpha|i\alpha\rangle+\langle \alpha|-i\alpha\rangle+\langle -\alpha|i\alpha\rangle+\langle -\alpha|-i\alpha\rangle\right)\\
&=\frac{(\alpha^2)^{4k}}{N_{+}}\left(\exp(-|\alpha|^{2})\exp(i|\alpha|^{2})+\exp(-|\alpha|^{2})\exp(-i|\alpha|^{2})+\exp(-|\alpha|^{2})\exp(-i|\alpha|^{2})\right.\\
&\left.+\exp(-|\alpha|^{2})\exp(i|\alpha|^{2})\right)\\
&=\frac{2(\alpha^2)^{4k}\exp(-|\alpha|^{2})}{N_{+}}\left(\exp(i|\alpha|^{2})+\exp(-i|\alpha|^{2})\right)\\
&=\frac{4(\alpha^2)^{4k}\exp(-|\alpha|^{2})}{N_{+}}\cos(|\alpha|^{2})\\
&=\frac{4(\alpha^2)^{4k}\exp(-|\alpha|^{2})}{4\exp(-|\alpha|^{2})\cosh(|\alpha|^{2})}\cos(|\alpha|^{2})=\frac{(\alpha^2)^{4k}\cos(|\alpha|^{2})}{\cosh(|\alpha|^{2})},
 \end{aligned}
\end{equation}

\begin{equation}
\begin{aligned}
\langle\bar{0}_{+}|(\hat{a}^{4k})^{\dagger}\hat{a}^{4k}|\bar{0}_{+}\rangle=\langle\bar{1}_{+}|(\hat{a}^{4k})^{\dagger}\hat{a}^{4k}|\bar{1}_{+}\rangle=(\alpha^2)^{4k},
\end{aligned}
\end{equation}

\begin{equation}
 \begin{aligned}
  \langle\bar{0}_{+}|(\hat{a}^{4k})^{\dagger}\hat{a}^{4k+1}|\bar{0}_{+}\rangle&=\langle\bar{0}_{+}|(\hat{a}^{4k})^{\dagger}\hat{a}^{4k+1}|\bar{1}_{+}\rangle\\
  =\langle\bar{1}_{+}|(\hat{a}^{4k})^{\dagger}\hat{a}^{4k+1}|\bar{1}_{+}\rangle&=\langle\bar{1}_{+}|(\hat{a}^{4k})^{\dagger}\hat{a}^{4k+1}|\bar{0}_{+}\rangle=0,
 \end{aligned}
\end{equation}

\begin{equation}
 \begin{aligned}
  \langle\bar{0}_{+}|(\hat{a}^{4k+1})^{\dagger}\hat{a}^{4k+1}|\bar{1}_{+}\rangle
  &=\frac{1}{N_{+}}(\langle \alpha|+\langle-\alpha|) (\hat{a}^{4k+1})^{\dagger}\hat{a}^{4k+1} ( |i\alpha\rangle+|-i\alpha\rangle)\\
  &=\frac{i (\alpha^{2})^{4k+1}}{N_{+}}(\langle \alpha|-\langle-\alpha|)( |i\alpha\rangle-|-i\alpha\rangle)\\
  &=\frac{i (\alpha^{2})^{4k+1}}{N_{+}}\left(\exp(-|\alpha|^{2})\exp(i|\alpha|^{2})-\exp(-|\alpha|^{2})\exp(-i|\alpha|^{2})\right.\\
  &\left.-\exp(-|\alpha|^{2})\exp(-i|\alpha|^{2})+\exp(-|\alpha|^{2})\exp(i|\alpha|^{2})\right)\\
 &=\frac{2i (\alpha^{2})^{4k+1}\exp(-\alpha^{2})}{N_{+}}\left(\exp(i|\alpha|^{2})-\exp(-i|\alpha|^{2})\right)\\
&=-\frac{4 (\alpha^{2})^{4k+1}\exp(-\alpha^{2})}{N_{+}}\sin(\alpha^{2})\\
&=-\frac{(\alpha^{2})^{4k+1}\sin(\alpha^{2})}{\cosh(\alpha^{2})},
 \end{aligned}
\end{equation}

\begin{equation}
 \begin{aligned}
  \langle\bar{0}_{+}|(\hat{a}^{4k+1})^{\dagger}\hat{a}^{4k+1}|\bar{0}_{+}\rangle
  &=\langle\bar{1}_{+}|(\hat{a}^{4k+1})^{\dagger}\hat{a}^{4k+1}|\bar{1}_{+}\rangle=\frac{(\alpha^{2})^{4k+1}N_{-}}{N_{+}},
\end{aligned}
\end{equation}
\noindent
Written in the Fock basis, the basic codewords read
\begin{equation}
 \begin{aligned}
  |\bar{0}_{+}\rangle&=\frac{1}{\sqrt{\cosh(\alpha^{2})}}\sum\limits_{n=0}^{\infty}\frac{\alpha^{2n}}{\sqrt{(2n)!}}|2n\rangle,\\
 |\bar{1}_{+}\rangle&=\frac{1}{\sqrt{\cosh(\alpha^{2}})}\sum\limits_{n=0}^{\infty}\frac{(-1)^{n}\alpha^{2n}}{\sqrt{(2n)!}}|2n\rangle.
 \end{aligned}
\end{equation}
\noindent
Like in the annihilation operator model, it is useful to study even and odd losses on the codewords separately:
\begin{equation} 
 \begin{aligned}
 A_{2m}|\bar{0}_{+}\rangle&=\frac{1}{\sqrt{\cosh(\alpha^{2})}}\sum\limits_{n=m}^{\infty}\frac{\alpha^{2n}}{\sqrt{(2n)!}}\sqrt{\gamma}^{2n-2m}\sqrt{1-\gamma}^{2m}\sqrt{\frac{(2n)!}{(2n-2m)!(2m)!}}|2n-2m\rangle\\
 &=\frac{1}{\sqrt{\cosh(\alpha^{2})}}\sum\limits_{n=m}^{\infty}\frac{\alpha^{2n}}{\sqrt{(2n-2m)!(2m)!}}\sqrt{\gamma}^{2n-2m}\sqrt{1-\gamma}^{2m}|2n-2m\rangle\\
 &=\frac{1}{\sqrt{(2m)!}\sqrt{\cosh(\alpha^{2})}}\sqrt{1-\gamma}^{2m}\alpha^{2m}\sum\limits_{l=0}^{\infty}\frac{\alpha^{2l}}{(2l)!}\sqrt{\gamma}^{2l}|2l\rangle\\
  &=\sqrt{\frac{\cosh(\alpha^{2}\gamma)}{\cosh(\alpha^{2})}}\frac{\sqrt{1-\gamma}^{2m}\alpha^{2m}}{\sqrt{(2m)!}}|\widetilde{0}_{+}\rangle\\
\end{aligned}
\end{equation}

\begin{equation}
\begin{aligned}
 A_{2m}|\bar{1}_{+}\rangle&=\frac{1}{\sqrt{\cosh(\alpha^{2})}}\sum\limits_{n=m}^{\infty}\frac{i^{2n}\alpha^{2n}}{\sqrt{(2n)!}}\sqrt{\gamma}^{2n-2m}\sqrt{1-\gamma}^{2m}\sqrt{\frac{(2n)!}{(2n-2m)!(2m)!}}|2n-2m\rangle\\
 &=\frac{1}{\sqrt{\cosh(\alpha^{2})}}\sum\limits_{n=m}^{\infty}\frac{i^{2n}\alpha^{2n}}{\sqrt{(2n-2m)!(2m)!}}\sqrt{\gamma}^{2n-2m}\sqrt{1-\gamma}^{2m}|2n-2m\rangle\\
 &=\frac{1}{\sqrt{(2m)!}\sqrt{\cosh(\alpha^{2})}}\sqrt{1-\gamma}^{2m}\alpha^{2m}i^{2m}\sum\limits_{l=0}^{\infty}(-1)^{l}\frac{\alpha^{2l}}{(2l)!}\sqrt{\gamma}^{2l}|2l\rangle\\
  &=\sqrt{\frac{\cosh(\alpha^{2}\gamma)}{\cosh(\alpha^{2})}}\frac{\sqrt{1-\gamma}^{2m}\alpha^{2m}i^{2m}}{\sqrt{(2m)!}}|\widetilde{1}_{+}\rangle\\
\end{aligned}
\end{equation}

\begin{equation}
\begin{aligned}
 A_{2m+1}|\bar{0}_{+}\rangle&=\frac{1}{\sqrt{\cosh(\alpha^{2})}}\sum\limits_{n=\Gamma}^{\infty}\frac{\alpha^{2n}}{\sqrt{(2n)!}}\sqrt{\gamma}^{2n-2m-1}\sqrt{1-\gamma}^{2m+1}\sqrt{\frac{(2n)!}{(2n-2m-1)!(2m+1)!}}|2n-2m-1\rangle\\
 &=\frac{1}{\sqrt{\cosh(\alpha^{2})}}\sum\limits_{n=\Gamma}^{\infty}\frac{\alpha^{2n}}{\sqrt{(2n-2m-1)!(2m+1)!}}\sqrt{\gamma}^{2n-2m-1}\sqrt{1-\gamma}^{2m+1}|2n-2m-1\rangle\\
 &=\frac{1}{\sqrt{(2m)!}\sqrt{\cosh(\alpha^{2})}}\sqrt{1-\gamma}^{2m+1}\alpha^{2m+1}\sum\limits_{l=0}^{\infty}\frac{\alpha^{2l+1}}{(2l+1)!}\sqrt{\gamma}^{2l+1}|2l+1\rangle\\
  &=\sqrt{\frac{\sinh(\alpha^{2}\gamma)}{\cosh(\alpha^{2})}}\frac{\sqrt{1-\gamma}^{2m+1}\alpha^{2m+1}}{\sqrt{(2m+1)!}}|\widetilde{0}_{-}\rangle\\
\end{aligned}
\end{equation}

\begin{equation}
\begin{aligned}
 A_{2m+1}|\bar{1}_{+}\rangle&=\frac{1}{\sqrt{\cosh(\alpha^{2})}}\sum\limits_{n=\Gamma}^{\infty}\frac{i^{2n}\alpha^{2n}}{\sqrt{(2n)!}}\sqrt{\gamma}^{2n-2m-1}\sqrt{1-\gamma}^{2m+1}\sqrt{\frac{(2n)!}{(2n-2m-1)!(2m+1)!}}|2n-2m-1\rangle\\
 &=\frac{1}{\sqrt{\cosh(\alpha^{2})}}\sum\limits_{n=m}^{\infty}\frac{i^{2n}\alpha^{2n}}{\sqrt{(2n-2m-1)!(2m+1)!}}\sqrt{\gamma}^{2n-2m-1}\sqrt{1-\gamma}^{2m+1}|2n-2m-1\rangle\\
 &=\frac{1}{\sqrt{(2m)!}\sqrt{\cosh(\alpha^{2})}}\sqrt{1-\gamma}^{2m+1}\alpha^{2m+1}i^{2m+1}\sum\limits_{l=0}^{\infty}(-1)^{l}\frac{\alpha^{2l+1}}{(2l+1)!}\sqrt{\gamma}^{2l}|2l+1\rangle\\
  &=\sqrt{\frac{\sinh(\alpha^{2}\gamma)}{\cosh(\alpha^{2})}}\frac{\sqrt{1-\gamma}^{2m+1}\alpha^{2m+1}i^{2m+1}}{\sqrt{(2m+1)!}}|\widetilde{1}_{-}\rangle\\
\end{aligned}
\end{equation}
One can easily verify that the norms of corrupted codewords are identical, i.e. the codewords are not deformed after loss. 
Qualitatively, we find the same cyclic behaviour as in the simplified loss model. Note however,
that the logical codewords in the different error spaces are not orthogonal for finite $\alpha$. The encoding presented here is
therefore not an exact QEC but can be regarded as an approximate QECC, provided $\alpha$ is taken sufficiently large to ensure near-orthogonality. \\  
From the expressions above, the probabilities for individual losses on the codewords can easily be determined by calculating the corresponding squared norm:
 \begin{equation}
 \begin{aligned}
 p_{0}&=\frac{\cosh(\alpha^{2}\gamma)}{\cosh(\alpha^{2})}\sum\limits_{m=0,2,4,}^{\infty}\frac{\left((1-\gamma)|\alpha|^{2}\right)^{2m}}{(2m)!}\\
 &=\frac{\cosh(\gamma \alpha^{2})}{2\cosh(\alpha^{2})}\left(\cos(-\alpha^{2}(1-\gamma))+\cosh(-\alpha^{2}(1-\gamma))\right)\\
 p_{1}&=\frac{\sinh(\alpha^{2}\gamma)}{\cosh(\alpha^{2})}\sum\limits_{m=0,2,4,}^{\infty}\frac{\left((1-\gamma)\alpha^{2}\right)^{2m+1}}{(2m+1)!}\\
 &=-\frac{\sinh(\gamma \alpha^{2})}{2\cosh(\alpha^{2})}\left(\sin(-\alpha^{2}(1-\gamma))+\sinh(-\alpha^{2}(1-\gamma))\right)\\
 p_{2}&=\frac{\cosh(\alpha^{2}\gamma)}{\cosh(\alpha^{2})}\sum\limits_{m=1,3,5,}^{\infty}\frac{\left((1-\gamma)\alpha^{2}\right)^{2m}}{(2m)!}\\
 &=\frac{\cosh(\gamma \alpha^{2})}{2\cosh(\alpha^{2})}\left(-\cos(-\alpha^{2}(1-\gamma))+\cosh(-\alpha^{2}(1-\gamma))\right)\\
 p_{3}&=\frac{\sinh(\alpha^{2}\gamma)}{\cosh(\alpha^{2})}\sum\limits_{m=1,3,5,}^{\infty}\frac{\left((1-\gamma)\alpha^{2}\right)^{2m+1}}{(2m+1)!}\\
&=-\frac{\sinh(\gamma \alpha^{2})}{2\cosh(\alpha^{2})}\left(\sin(-\alpha^{2}(1-\gamma))-\sinh(-\alpha^{2}(1-\gamma))\right)
 \end{aligned}
 \end{equation}
 Note that $p_{0}$ contains the probability for no loss, four losses, eight losses and so on. \\
 We are interested in the evolution of a logical qubit subject to photon loss. 
 Taking the finite overlap of the codewords into account, a properly normalised qubit reads as
 \begin{equation}
  |\bar{\psi}\rangle=\frac{a|\bar{0}_{+}\rangle+b|\bar{1}_{+}\rangle}{\sqrt{1+2\operatorname{Re}(ab^{*})\langle\bar{0}_{+}|\bar{1}_{+}\rangle}}.
 \end{equation}
The codewords are not deformed, such that after a loss, say one loss and the cyclic equivalents, a global factor $p_{1}$ arises. In addition to that, we have to take the 
non-orthogonality of the codewords in the error spaces into account, i.e. we have to properly renormalise the erroneous state. In this example, we
get a relative phase of $i$ which changes the norm of the qubit as well as the damped amplitude:
\begin{equation}
 A_{1}|\bar{\psi}\rangle=\sqrt{\frac{1-2\operatorname{Re}(a^{*}b)\frac{\sin(\gamma\alpha^{2})}{\sinh(\gamma\alpha^{2})}}{1+2\operatorname{Re}(ab^{*})\frac{\cos(\alpha^{2})}{\cosh(\alpha^{2})}}p_{1}} \left(\frac{a|\widetilde{0}_{-}\rangle+ib|\widetilde{1}_{-}\rangle}{\sqrt{1-2\operatorname{Re}(a^{*}b)\frac{\sin(\gamma\alpha^{2})}{\sinh(\gamma\alpha^{2})}}}\right).\\
\end{equation}
The loss probability is therefore
\begin{equation}
\widetilde{p_{1}}=\frac{1-2\operatorname{Re}(a^{*}b)\frac{\sin(\gamma\alpha^{2})}{\sinh(\gamma\alpha^{2})}}{1+2\operatorname{Re}(ab^{*})\frac{\cos(\alpha^{2})}{\cosh(\alpha^{2})}}p_{1}.
\end{equation}
and the normalised state reads
\begin{equation}
 \left(\frac{a|\widetilde{0}_{-}\rangle+ib|\widetilde{1}_{-}\rangle}{\sqrt{1-2\operatorname{Re}(a^{*}b)\frac{\sin(\gamma\alpha^{2})}{\sinh(\gamma\alpha^{2})}}}\right).
\end{equation}

The analogous results for the loss probabilities and the total final mixed states are summarised in the main text.\\
For comparison, let us investigate the KL conditions for the codewords in the $\bar{X}$-basis (see also the discussion after Eq. \eqref{eq: xbasis} in the main text):
\begin{equation}
 \begin{aligned}
  |\bar{0}_{+}+\bar{1}_{+}\rangle&=\frac{1}{\sqrt{N_{+}^{\prime}}}(|\bar{0}_{+}\rangle+|\bar{1}_{+}\rangle),\\
  |\bar{0}_{+}-\bar{1}_{+}\rangle&=\frac{1}{\sqrt{N_{-}^{\prime}}}(|\bar{0}_{+}\rangle-|\bar{1}_{+}\rangle).
\end{aligned}
\end{equation}
We consider again the error set $\mathcal{E}=\{\hat{a}^{4k},\hat{a}^{4k+1}, k\in \mathbb{N}_{0}\}$. The action on the $\bar{X}$-basis codewords can be easily calculated
based on the results of the $\bar{Z}$-basis analysis:
\begin{equation}
 \begin{aligned}
  \hat{a}^{4k}|\bar{0}_{+}+\bar{1}_{+}\rangle&=\alpha^{4k}|\bar{0}_{+}+\bar{1}_{+}\rangle=\frac{\alpha^{4k}}{\sqrt{N_{+}^{\prime}}}(|\bar{0}_{+}\rangle+|\bar{1}_{+}\rangle),\\
  \hat{a}^{4k}|\bar{0}_{+}-\bar{1}_{+}\rangle&=\alpha^{4k}|\bar{0}_{+}-\bar{1}_{+}\rangle=\frac{\alpha^{4k}}{\sqrt{N_{-}^{\prime}}}(|\bar{0}_{+}\rangle-|\bar{1}_{+}\rangle),\\
 \hat{a}^{4k+1}|\bar{0}_{+}+\bar{1}_{+}\rangle&=\frac{\alpha^{4k+1}}{\sqrt{N_{+}^{\prime}}}(|\bar{0}_{-}\rangle+i|\bar{1}_{-}\rangle),\\
 \hat{a}^{4k+1}|\bar{0}_{+}+\bar{1}_{+}\rangle&=\frac{\alpha^{4k+1}}{\sqrt{N_{-}^{\prime}}}(|\bar{0}_{-}\rangle-i|\bar{1}_{-}\rangle).\\
 \end{aligned}
\end{equation}

The orthogonality requirements are fulfilled, as
\begin{equation}
 \begin{aligned}
  \langle\bar{0}_{+}+\bar{1}_{+}|(\hat{a}^{4k})^{\dagger}\hat{a}^{4k}|\bar{0}_{+}-\bar{1}_{+}\rangle&=\langle\bar{0}_{+}+\bar{1}_{+}|(\hat{a}^{4k+1})^{\dagger}\hat{a}^{4k+1}|\bar{0}_{+}-\bar{1}_{+}\rangle\\
 &=\langle\bar{0}_{+}+\bar{1}_{+}|(\hat{a}^{4k+1})^{\dagger}\hat{a}^{4k}|\bar{0}_{+}-\bar{1}_{+}\rangle\\
 &=\langle\bar{0}_{+}-\bar{1}_{+}|(\hat{a}^{4k+1})^{\dagger}\hat{a}^{4k}|\bar{0}_{+}+\bar{1}_{+}\rangle.\\
 \end{aligned}
\end{equation}

The non-deformation criterion for $\hat{a}^{4k}$ reads
\begin{equation}
 \begin{aligned}
  \langle\bar{0}_{+}+\bar{1}_{+}|(\hat{a}^{4k})^{\dagger}\hat{a}^{4k}|\bar{0}_{+}-\bar{1}_{+}\rangle&=\langle\bar{0}_{+}-\bar{1}_{+}|(\hat{a}^{4k})^{\dagger}\hat{a}^{4k}|\bar{0}_{+}-\bar{1}_{+}\rangle=(\alpha^{2})^{4k}.
 \end{aligned}
\end{equation}
However, for $\hat{a}^{4k+1}$, we have
\begin{equation}
 \begin{aligned}
  \langle\bar{0}_{+}+\bar{1}_{+}|(\hat{a}^{4k+1})^{\dagger}\hat{a}^{4k+1}|\bar{0}_{+}+\bar{1}_{+}\rangle&=\frac{2(\alpha^{2})^{4k+1}}{N_{+}^{\prime}}\left(1-\frac{\sin(\alpha^{2})}{\sinh(\alpha^{2})}\right),\\
 \langle\bar{0}_{+}-\bar{1}_{+}|(\hat{a}^{4k+1})^{\dagger}\hat{a}^{4k+1}|\bar{0}_{+}-\bar{1}_{+}\rangle&=\frac{2(\alpha^{2})^{4k+1}}{N_{-}^{\prime}}\left(1+\frac{\sin(\alpha^{2})}{\sinh(\alpha^{2})}\right),
 \end{aligned}
\end{equation}
which shows a violation of the non-deformation criterion. This can only be overcome with sufficiently large $\alpha$.

\section{Error correction steps and lower bound on fidelity}
\label{sec: fidelity}

We discuss the error correction procedure and the fidelity as a figure of merit for the example of the one-loss cat code. An extension to the
higher-loss codes is straight-forward.\\
The first step for correcting loss-induced errors on an incoming logical qubit is to determine its photon number parity (even or odd) and hence the subspace
in which the qubit resides (code or error space). After this first error-correction step, i.e., the number parity measurement that projects $\bar{\rho}$ either onto the even space with the normalized conditional density matrix
 \begin{equation}
  \begin{aligned}
   \rho^{(+)}&=\frac{1}{P^{{+}}}\left[\widetilde{p}_{0} \left(\frac{a|\widetilde{0}_{+}\rangle+b|\widetilde{1}_{+}\rangle}{\sqrt{1+2\operatorname{Re}(ab^{*})\frac{\cos(\gamma \alpha^{2})}{\cosh(\gamma \alpha^{2})}}}\right)\times H.c.\right.\\
 &\left.+\widetilde{p}_{2} \left(\frac{a|\widetilde{0}_{+}\rangle-b|\widetilde{1}_{+}\rangle}{\sqrt{1-2\operatorname{Re}(ab^{*})\frac{\cos(\gamma\alpha^{2})}{\cosh(\gamma\alpha^{2})}}}\right)\times H.c.\right],\\
  \end{aligned}
\end{equation}

or onto the odd space with the corresponding density matrix 

\begin{equation}
  \begin{aligned}
   \rho^{(-)}&=\frac{1}{P^{{-}}}\left[\widetilde{p}_{1} \left(\frac{a|\widetilde{0}_{-}\rangle+ib|\widetilde{1}_{-}\rangle}{\sqrt{1-2\operatorname{Re}(ab^{*})\frac{\sin(\gamma \alpha^{2})}{\sinh(\gamma \alpha^{2})}}}\right)\times H.c.\right.\\
 &\left.+\widetilde{p}_{3} \left(\frac{a|\widetilde{0}_{-}\rangle-ib|\widetilde{1}_{-}\rangle}{\sqrt{1+2\operatorname{Re}(ab^{*})\frac{\sin(\gamma\alpha^{2})}{\sinh(\gamma\alpha^{2})}}}\right)\times H.c.\right],\\
  \end{aligned}
\end{equation}

as a second step, the amplitudes are probabilistically restored, $|\widetilde{0}_{+}\rangle\rightarrow |\bar{0}_{+}\rangle$, etc. (see App. \ref{sec: restauration}).
Here, $P^{{+}}$ and  $P^{{-}}$ are the probabilities for obtaining the error syndromes "even" (code space) and "odd" (error space), respectively (they correspond
to the trace of the respective expression in squared brackets, i.e. each unnormalised conditional state). The worst-case fidelity is defined
as
\begin{equation}
\label{eq: worstcase}
 \begin{aligned}
  F_{wc}&=\min_{a,b}\left[\langle\bar{\psi}|\hat{\rho}^{(+)}|\bar{\psi}\rangle P^{{+}}+ \langle\bar{\psi}^{\prime}|\hat{\rho}^{(-)}|\bar{\psi}^{\prime}\rangle P^{{-}}\right]\\
 &=\min_{a,b}\left[\widetilde{p}_{0}(a,b)+\widetilde{p}_{2}(a,b)\Bigl\lvert\langle\bar{\psi}|\left(\frac{a|\bar{0}_{+}\rangle-b|\bar{1}_{+}\rangle}{\sqrt{1-2\operatorname{Re}(ab^{*})\frac{\cos(\alpha^{2})}{\cosh(\alpha^{2})}}}\right)\Bigr\rvert^{2}\right.\\
 &\left.+\widetilde{p}_{1}(a,b)+\widetilde{p}_{3}(a,b)\Bigl\lvert\langle\bar{\psi}^{\prime}|\left(\frac{a|\bar{0}_{+}\rangle-ib|\bar{1}_{+}\rangle}{\sqrt{1-2\operatorname{Re}(iab^{*})\frac{\cos(\alpha^{2})}{\cosh(\alpha^{2})}}}\right)\Bigr\rvert^{2}\right],
 \end{aligned}
\end{equation}
where $|\bar{\psi}^{\prime}\rangle=\frac{a|\bar{0}_{+}\rangle+ib|\bar{1}_{+}\rangle}{\sqrt{1-2\operatorname{Re}(iab^{*})\frac{\cos(\alpha^{2})}{\cosh(\alpha^{2})}}}$ 
(i.e., the reference input state for the odd syndrome has
a fixed phase gate applied to it compared to the original qubit input state 
$|\bar{\psi}\rangle=\frac{a|\bar{0}_{+}\rangle+b|\bar{1}_{+}\rangle}{\sqrt{1+2\operatorname{Re}(ab^{*})\frac{\cos(\alpha^{2})}{\cosh(\alpha^{2})}}}$).\\

The second term in each of the last two lines of Eq. \eqref{eq: worstcase} is non-negative. If both terms vanish (i.e., we have $\alpha\rightarrow \infty$ and
$a=\frac{1}{\sqrt{2}}=\pm b$), $\widetilde{p}_{0}$ and $\widetilde{p}_{1}$ no longer depend on $a$ and $b$, and $F_{wc}=\widetilde{p}_{0}+\widetilde{p}_{1}$
(more generally: $F_{wc}= \widetilde{p}_{0}(a^{wc},b^{wc})+\widetilde{p}_{1}(a^{wc},b^{wc})\geq \min\limits_{a,b}~ (\widetilde{p}_{0}+\widetilde{p}_{1})$).\\
For the other case when the two relevant terms in Eq. \eqref{eq: worstcase} do not vanish, we have
\begin{equation}
\begin{aligned}
 F_{wc}&>\widetilde{p}_{0}(a^{wc},b^{wc})+\widetilde{p}_{1}(a^{wc},b^{wc})\\
 &\geq\min\limits_{a,b}~ (\widetilde{p}_{0}+\widetilde{p}_{1})
\end{aligned}
\end{equation}
Thus, in general, we obtain the bound on $F_{wc}$ as expressed by Eq. \eqref{eq: bound}.\\
We show in the following that the state-dependent fidelity $F(a,b)$ for $L=1$ and a logical qubit $a|\bar{0}_{+}\rangle+b|\bar{1}_{+}\rangle$ 
under the conditions $a,b \in \mathbb{R}$ is extremal for  $|a|=|b|=\frac{1}{\sqrt{2}}$. \\
The fidelity can be cast in the following form:
 \begin{equation}
 \begin{aligned}
  F(a,b)&=\frac{1+2ab c_{1}}{1+2ab c_{2}}p_{0}+\frac{1-2ab c_{3}}{1+2ab c_{2}}p_{1}\\
  &=\frac{1+2a\sqrt{1-a^{2}} c_{1}}{1+2a\sqrt{1-a^{2}} c_{2}}p_{0}+\frac{1-2a\sqrt{1-a^{2}} c_{3}}{1+2a\sqrt{1-a^{2}} c_{2}}p_{1}\\
  &=\frac{p_{0}+p_{1}+2a\sqrt{1-a^{2}}(c_{1}p_{0}-c_{3}p_{1})}{1+2a\sqrt{1-a^{2}} c_{2}}\\
  &=F(a).
 \end{aligned}
 \end{equation}
 Here, the coefficients are short-hand for the overlaps of the codewords in the different error spaces, see Eqs. (13-15), and Eqs. \eqref{eq: plusoverlap} and \eqref{eq: minusoverlap}. These coefficients are real and 
 bounded by 1. To find the extremal value of the fidelity, we derive $F$ with respect to $a$:
\begin{equation}
 \begin{aligned}
 \frac{dF}{da}&=\frac{(1+2a\sqrt{1-a^{2}} c_{2})(c_{1}p_{0}-c_{3}p_{1})\frac{2-4a^{2}}{\sqrt{1-a^{2}}}}{(1+2a\sqrt{1-a^{2}} c_{2})^{2}}\\
& -\frac{(p_{0}+p_{1}+2a\sqrt{1-a^{2}}(c_{1}p_{0}-c_{3}p_{1}))c_{2}\frac{2-4a^{2}}{\sqrt{1-a^{2}}}}{(1+2a\sqrt{1-a^{2}} c_{2})^{2}}\\
&\propto 2-4a^{2}.
\end{aligned}
\end{equation}
This vanishes for $a^{2}=\frac{1}{2}$ and therefore  we find the two solutions $a=\pm\frac{1}{\sqrt{2}}$. One solution corresponds
to a maximum and the other to a minimum. The second derivative can resolve this and the solution depends on the signs of the coefficients $c_{i}$. 
To be safe and to avoid complicated formulae, one can clearly set
\begin{equation}
f_{min}(a):=\min \{F(a=\frac{1}{\sqrt{2}}), F(a=-\frac{1}{\sqrt{2}})\},
\end{equation}
which then corresponds to the lower bound $F$ on the worst-case fidelity $F_{wc}$, as shown in Eq. \eqref{eq: bound}.

\newpage
\section{Derivation of codewords in the Fock basis}
\label{sec: derivation}
To solve the system of equations \eqref{eq: definition}, we set $|\bar{0}\rangle=\sum\limits_{n=0}^{\infty}c_{n}|n\rangle$. We have
\begin{equation}
\begin{aligned}
 \hat{a}^{L+1}|\bar{0}\rangle&=\sum\limits_{n=L+1}^{\infty}c_{n}\sqrt{n}\sqrt{n-1}\cdots \sqrt{n-L}|n-L-1\rangle\\
 &=\sum\limits_{k=0}^{\infty}c_{k+L+1}\sqrt{k+L+1}\sqrt{k+L}\cdots \sqrt{k+1}|k\rangle\\
 &=\alpha^{L+1}\sum\limits_{k=0}^{\infty}c_{k}|k\rangle
\end{aligned}
\end{equation}
One obtains a recursive definition of the coefficients:
\begin{equation}
 c_{k+L+1}=\frac{\alpha^{L+1}c_{k}}{\sqrt{k+L+1}\sqrt{k+L}\cdots\sqrt{k+1}}
\end{equation}
\noindent
To solve the series, one of the first parameters, $c_{0}$ or $c_{1}$, has to be fixed.\\
Before determining the general solution, let us examine the easiest example, $L=0$. Here, the parity condition is trivial and
the other two equations are just the defining equations for coherent states. Therefore,the system of equations leads
to $|\bar{0}\rangle=|\alpha\rangle$ and $|\bar{1}\rangle=|-\alpha\rangle$.\\
As another illustrative example, we choose $L=1$. Then we find the series:
\begin{equation}
 c_{k+2}=\frac{\alpha^{2}c_{k}}{\sqrt{k+2}\sqrt{k+1}}
\end{equation}
\noindent
If we set $c_{0}$ as given, this series is resolved by
\begin{equation}
\label{eq: recursion}
c_{2k}=\frac{\alpha^{2k} c_{0}}{\sqrt{(2k)!}}.
\end{equation}
However, $c_{0}$ is not arbitrary because it follows from the normalisation constraint
\begin{equation}
 \sum\limits_{k=0}^{\infty}|c_{2k}|^{2}=\sum\limits_{k=0}^{\infty}\frac{(\alpha^{2})^{2k}}{(2k)!}|c_{0}|^{2}=1\Rightarrow |c_{0}|
\end{equation}
\noindent
The coefficient $c_{0}$ is then determined up to a irrelevant phase which leads to a global phase because of \eqref{eq: recursion}.
The corresponding recursion formula for $|\bar{1}\rangle$ is given by
\begin{equation}
c_{2k}=\frac{(-1)^{k}\alpha^{2k} c_{0}}{\sqrt{(2k)!}},
\end{equation}
where $c_{0}$ can be determined through normalisation. Therefore, in the Fock basis, the basis codewords read:
\begin{equation}
\begin{aligned}
 |\bar{0}\rangle&=\frac{1}{\sqrt{\cosh(\alpha^{2})}}\sum\limits_{n=0}^{\infty}\frac{\alpha^{2n}}{\sqrt{(2n)!}}|2n\rangle,\\
 |\bar{1}\rangle&=\frac{1}{\sqrt{\cosh(\alpha^{2})}}\sum\limits_{n=0}^{\infty}(-1)^{n}\frac{\alpha^{2n}}{\sqrt{(2n)!}}|2n\rangle.\\
 \end{aligned}
 \end{equation}
This can be expressed in terms of coherent states
 \begin{equation}
 \begin{aligned}
 |\bar{0}\rangle&=\frac{1}{\sqrt{N_{+}}}(|\alpha\rangle+|-\alpha\rangle)\\
 |\bar{1}\rangle&=\frac{1}{\sqrt{N_{+}}}(|i\alpha\rangle+|-i\alpha\rangle).\\
 \end{aligned}
 \end{equation}
 as given in Sec. \ref{sec: cat-code}.\\
 If we fix $c_{1}$, a completely analogue calculation leads to
 \begin{equation}
 \begin{aligned}
 |\bar{0}\rangle&=\frac{1}{\sqrt{\sinh(\alpha^{2})}}\sum\limits_{n=0}^{\infty}\frac{\alpha^{2n+1}}{\sqrt{(2n+1)!}}|2n+1\rangle\\
 |\bar{1}\rangle&=\frac{1}{\sqrt{\sinh(\alpha^{2})}}\sum\limits_{n=0}^{\infty}(-1)^{n}\frac{\alpha^{2n+1}}{\sqrt{(2n+1)!}}|2n+1\rangle.\\
 \end{aligned}
 \end{equation}
 Since the parity condition for $L=1$ reads as $(-1)^{\hat{n}}|\bar{\psi}\rangle=|\bar{\psi}\rangle$, the first pair of codewords is
 the solution of the determining system of equations.\\
 For general $L$, it is easy to verify that the solutions of the defining equations in the Fock basis are given by
 \begin{equation}
 \begin{aligned}
 |\bar{0}\rangle&=\sum\limits_{k=0}^{\infty}\frac{\alpha^{(L+1)k}}{\sqrt{([L+1]k)!}}|(L+1)k\rangle,\\
 |\bar{1}\rangle&=\sum\limits_{k=0}^{\infty}\frac{(-1)^{k}\alpha^{(L+1)k}}{\sqrt{([L+1]k)!}}|(L+1)k\rangle.\\
 \end{aligned}
 \end{equation}
 We show in the following these states can be rewritten in terms of coherent states as presented in the main text:
  \begin{equation}
  \begin{aligned}
  |\bar{0}\rangle&=\sum\limits_{k=0}^{L}|\alpha \exp\left(\frac{2\pi i k}{L+1}\right)\rangle,\\
 |\bar{1}\rangle&=\sum\limits_{k=1}^{L+1}|\alpha \exp\left(\frac{(2k-1)\pi i }{L+1}\right)\rangle.\\
 \end{aligned}
 \end{equation}
 Expressed in the Fock basis, we have

 \begin{align*}
  |\bar{0}\rangle&=\sum\limits_{k=0}^{L}\sum\limits_{r=0}^{\infty}\frac{\alpha^{r}\exp\left(\frac{\pi i 2k}{L+1}\right)^{r}}{\sqrt{r!}}|r\rangle \numberthis\\
  &=\sum\limits_{r=0}^{\infty}\frac{\alpha^{r}}{\sqrt{r!}}|r\rangle\sum\limits_{k=0}^{L}\exp\left(\frac{\pi i 2r}{L+1}\right)^{k}\\
  &=\sum\limits_{r=0}^{\infty}\frac{\alpha^{r}}{\sqrt{r!}}|r\rangle\frac{1-\exp\left(\pi i 2r\right)}{1-\exp\left(\frac{\pi i 2r}{L+1}\right)}\\
 &=\sum\limits_{r=0}^{\infty}\frac{\alpha^{r}}{\sqrt{r!}}|r\rangle \delta_{r}^{(L+1)m}\\
 &=\sum\limits_{m=0}^{\infty}\frac{\alpha^{(L+1)m}}{\sqrt{[(L+1)m]!}}|(L+1)m\rangle
 \end{align*}

The calculation for the other codeword is similar,
 \begin{align*}
 |\bar{1}\rangle&=\sum\limits_{k=1}^{L+1}\sum\limits_{r=0}^{\infty} \frac{\alpha^{r}\exp\left(\frac{\pi i (2k-1)}{L+1}\right)^{r}}{\sqrt{r!}}|r\rangle\numberthis\\
  \displaybreak
 &=\sum\limits_{r=0}^{\infty}\frac{\alpha^{r}}{\sqrt{r!}}|r\rangle\sum\limits_{k=1}^{L+1}\exp\left(\frac{\pi i r}{L+1}\right)^{2k-1}\\
 &=\sum\limits_{r=0}^{\infty}\frac{\alpha^{r}}{\sqrt{r!}}|r\rangle\exp\left(-\frac{\pi i r}{L+1}\right)\sum\limits_{k=1}^{L+1}\exp\left(\frac{2\pi i r}{L+1}\right)^{k}\\
 &=\sum\limits_{r=0}^{\infty}\frac{\alpha^{r}}{\sqrt{r!}}|r\rangle\exp\left(\frac{\pi i r}{L+1}\right)\sum\limits_{j=0}^{L}\exp\left(\frac{2\pi i r}{L+1}\right)^{j}\\
 &=\sum\limits_{r=0}^{\infty}\frac{\alpha^{r}}{\sqrt{r!}}|r\rangle\exp\left(\frac{\pi i r}{L+1}\right)\frac{1-\exp\left(\pi i 2r\right)}{1-\exp\left(\frac{\pi i 2r}{L+1}\right)}\\
 &=\sum\limits_{r=0}^{\infty}\frac{\alpha^{r}}{\sqrt{r!}}|r\rangle\exp\left(\frac{\pi i r}{L+1}\right) \delta_{r}^{(L+1)m}\\
 &=\sum\limits_{m=0}^{\infty}\frac{(-1)^{m}\alpha^{(L+1)m}}{\sqrt{[(L+1)m]!}}|(L+1)m\rangle.\\
 \end{align*}

Up to normalisation, the basic codewords in the $L+1$ error spaces in terms of coherent states are defined as
\begin{equation}
\begin{aligned}
|\bar{0}_{q}\rangle_{L}:&=\sum\limits_{k=0}^{L}\exp\left(\frac{2q ki\pi}{L+1}\right)|\alpha \exp\left(\frac{2 k i\pi}{L+1}\right)\rangle,\\
|\bar{1}_{q}\rangle_{L}:&=\sum\limits_{k=1}^{L+1}\exp\left(\frac{2q(k-1)i\pi}{L+1}\right)|\alpha \exp\left(\frac{(2k-1)i\pi}{L+1}\right)\rangle,
\end{aligned}
\end{equation}
for $q=0,.., L$. In the Fock basis, these can be expressed as

\begin{equation}
\begin{aligned}
 |\bar{0}_{q}\rangle_{L}:=\sum\limits_{k=1}^{\infty}\frac{\alpha^{(L+1)k-q}}{\sqrt{[(L+1)k-q]!}}|(L+1)k-q\rangle,\\
 |\bar{1}_{q}\rangle_{L}:=\sum\limits_{k=1}^{\infty}\frac{(e^{\frac{i\pi}{L+1}}\alpha)^{(L+1)k-q}}{\sqrt{[(L+1)k-q]!}}|(L+1)k-q\rangle.\\
\end{aligned}
\end{equation}
The code-defining equations, including both the code space and all error spaces, then become

 \begin{align*}
  \exp\left(\frac{2\pi i \hat{n}}{L+1}\right)|\bar{0}_{q}\rangle_{L}&=\exp\left(\frac{2\pi i q}{L+1}\right)|\bar{0}_{q}\rangle_{L}, \numberthis\\
   \exp\left(\frac{2\pi i \hat{n}}{L+1}\right)|\bar{1}_{q}\rangle_{L}&=\exp\left(\frac{2\pi i q}{L+1}\right)|\bar{1}_{q}\rangle_{L},\\
   (\hat{a}^{L+1}-\alpha^{L+1})|\bar{0}_{q}\rangle_{L}&=0,\\
    (\hat{a}^{L+1}+\alpha^{L+1})|\bar{1}_{q}\rangle_{L}&=0,
 \end{align*}

$\forall q=0,1,2,...,L$. The evolution of a logical qubit $|\bar{\Psi}\rangle=\frac{a|\bar{0}\rangle+b|\bar{1}\rangle}{\sqrt{1+2\operatorname{Re}(ab^{*}\langle\bar{0}|\bar{1}\rangle)}}$
under AD can then be described as an (unnormalised) mixture of $2(L+1)$ components:
\begin{align*}
 \bar{\rho} &= p_{0}(a|\widetilde{0}_{0}\rangle+b|\widetilde{1}_{0}\rangle)\times H.c. \numberthis
 +p_{1}(a|\widetilde{0}_{1}\rangle+e^{\frac{i \pi}{L+1}} b|\widetilde{1}_{1}\rangle)\times H.c.
+p_{2}(a|\widetilde{0}_{2}\rangle+e^{\frac{2i \pi}{L+1}} b|\widetilde{1}_{2}\rangle)\times H.c.\\
 &+p_{3}(a|\widetilde{0}_{3}\rangle+e^{\frac{3i \pi}{L+1}} b|\widetilde{1}_{3}\rangle)\times H.c.
 +...
 +p_{L}(a|\widetilde{0}_{L}\rangle+e^{\frac{Li \pi}{L+1}} b|\widetilde{1}_{L}\rangle)\times H.c.\\
 &+p_{L+1}(a|\widetilde{0}_{0}\rangle_{1}- b|\widetilde{1}_{0}\rangle_{1})\times H.c.
 +p_{L+2}(a|\widetilde{0}_{1}\rangle-e^{\frac{i \pi}{L+1}} b|\widetilde{1}_{1}\rangle)\times H.c.\\
 &+p_{L+3}(a|\widetilde{0}_{2}\rangle-e^{\frac{2i \pi}{L+1}} b|\widetilde{1}_{2}\rangle_{2})\times H.c.
 +...+p_{2L+1}(a|\bar{0}_{L}\rangle-e^{\frac{Li \pi}{L+1}} b|\widetilde{1}_{L}\rangle)\times H.c.\\
\end{align*}

\newpage
\section{Full loss channel and KL conditions for the two-loss code}
\label{sec: twoloss}
The normalised $L=2$ codewords read in the Fock basis as:
\begin{equation}
\begin{aligned}
 |\bar{0}\rangle&=\frac{1}{\sqrt{N}}\sum\limits_{k=0}^{\infty}\frac{\alpha^{3k}}{\sqrt{(3k)!}}|3k\rangle,\\
 |\bar{1}\rangle&=\frac{1}{\sqrt{N}}\sum\limits_{k=0}^{\infty}\frac{(-\alpha)^{3k}}{\sqrt{(3k)!}}|3k\rangle,
\end{aligned}
\end{equation}
where $N=\frac{1}{3}\left(\exp(\alpha^{2})+2\exp\left(\frac{-\alpha^{2}}{2}\right)\cos\left(\frac{\sqrt{3}\alpha^{2}}{2}\right)\right)$. The codeword
overlap is given by 
\begin{equation}
 \langle\bar{1}|\bar{0}\rangle=\frac{\exp(-\alpha^{2})+2\exp\left(\frac{\alpha^{2}}{2}\right)\cos\left(\frac{\sqrt{3}\alpha^{2}}{2}\right)}{\exp(\alpha^{2})+2\exp\left(\frac{-\alpha^{2}}{2}\right)\cos\left(\frac{\sqrt{3}\alpha^{2}}{2}\right)}
\xrightarrow[\alpha \to \infty]{} 0.
 \end{equation}

Based on the results obtained in Sec. \ref{sec: generalised cat codes} for the simplified error model, we expect a similar cyclic behaviour of 
the code under the real AD channel. It is therefore advantageous to consider the action of the operators $\{A_{3k},A_{3k+1}, A_{3k+2}\}$
on the codewords:
\begin{equation}
\label{eq: L2AD}
\begin{aligned}
 A_{3k}|\bar{0}\rangle&=\frac{1}{\sqrt{N}}\frac{\sqrt{1-\gamma}^{3k}\alpha^{3k}}{\sqrt{(3k)!}}\sum\limits_{n=0}^{\infty}\frac{(\alpha\sqrt{\gamma})^{3n}}{\sqrt{(3n)!}}|3n\rangle,\\
 A_{3k}|\bar{1}\rangle&=\frac{(-1)^{k}}{\sqrt{N}}\frac{\sqrt{1-\gamma}^{3k}\alpha^{3k}}{\sqrt{(3k)!}}\sum\limits_{n=0}^{\infty}\frac{(-\alpha\sqrt{\gamma})^{3n}}{\sqrt{(3n)!}}|3n\rangle,\\
 A_{3k+1}|\bar{0}\rangle&=\frac{1}{\sqrt{N}}\frac{\sqrt{1-\gamma}^{3k+1}\alpha^{3k+1}}{\sqrt{(3k+1)!}}\sum\limits_{n=1}^{\infty}\frac{(\alpha\sqrt{\gamma})^{3n-1}}{\sqrt{(3n-1)!}}|3n-1\rangle,\\
 A_{3k+1}|\bar{1}\rangle&=\frac{(-1)^{k+1}}{\sqrt{N}}\frac{\sqrt{1-\gamma}^{3k+1}\alpha^{3k+1}e^{\frac{\pi i}{3}}}{\sqrt{(3k+1)!}}\sum\limits_{n=1}^{\infty}\frac{(e^{\frac{\pi i}{3}}\alpha\sqrt{\gamma})^{3n-1}}{\sqrt{(3n-1)!}}|3n-1\rangle,\\
 A_{3k+2}|\bar{0}\rangle&=\frac{1}{\sqrt{N}}\frac{\sqrt{1-\gamma}^{3k+2}\alpha^{3k+2}}{\sqrt{(3k+2)!}}\sum\limits_{n=1}^{\infty}\frac{(\alpha\sqrt{\gamma})^{3n-2}}{\sqrt{(3n-2)!}}|3n-2\rangle,\\
 A_{3k+2}|\bar{1}\rangle&=\frac{(-1)^{k}}{\sqrt{N}}\frac{\sqrt{1-\gamma}^{3k+2}\alpha^{3k+2}e^{\frac{2 \pi i}{3}}}{\sqrt{(3k+2)!}}\sum\limits_{n=1}^{\infty}\frac{(e^{\frac{ \pi i}{3}}\alpha\sqrt{\gamma})^{3n-2}}{\sqrt{(3n-2)!}}|3n-2\rangle.\\
 \end{aligned}
 \end{equation}
Following the notation introduced after Eq. \eqref{eq: L2notation}, the basic codewords in the three orthogonal error spaces read  
\begin{equation}
\begin{aligned}
 |\widetilde{0}_{0}\rangle_{2}&\propto\sum\limits_{k=0}^{\infty}\frac{(\alpha\sqrt{\gamma})^{3k}}{\sqrt{(3k)!}}|3k\rangle,\\
 |\widetilde{1}_{0}\rangle_{2}&\propto\sum\limits_{k=0}^{\infty}\frac{(-\alpha\sqrt{\gamma})^{3k}}{\sqrt{(3k)!}}|3k\rangle,\\
 |\widetilde{0}_{1}\rangle_{2}&\propto\sum\limits_{k=1}^{\infty}\frac{(\alpha\sqrt{\gamma})^{3k-1}}{\sqrt{(3k-1)!}}|3k-1\rangle,\\
 |\widetilde{1}_{1}\rangle_{2}&\propto\sum\limits_{k=1}^{\infty}\frac{(e^{\frac{\pi i}{3}}\alpha\sqrt{\gamma})^{3k-1}}{\sqrt{(3k-1)!}}|3k-1\rangle,\\
 |\widetilde{0}_{2}\rangle_{2}&\propto\sum\limits_{k=1}^{\infty}\frac{(\alpha\sqrt{\gamma})^{3k-2}}{\sqrt{(3k-2)!}}|3k-2\rangle,\\
 |\widetilde{1}_{2}\rangle_{2}&\propto\sum\limits_{k=1}^{\infty}\frac{(e^{\frac{\pi i}{3}}\alpha\sqrt{\gamma})^{3k-2}}{\sqrt{(3k-2)!}}|3k-2\rangle,
 \end{aligned}
\end{equation}
where $\sim$ again indicates the damped amplitude. Obviously, the different error spaces are orthogonal and the codewords in 
each error space become orthogonal for large $\alpha$.\\
Furthermore, we define the following logical states ($\sim$ denotes again damped logical states)
\begin{equation}
\begin{aligned}
 |\widetilde{\Psi}_{0}\rangle&\propto a|\widetilde{0}_{0}\rangle_{2}+b|\widetilde{1}_{0}\rangle_{2},\\
 |\widetilde{\Psi}_{1}\rangle&\propto a|\widetilde{0}_{1}\rangle_{2}+e^{\frac{\pi i}{3}}b|\widetilde{1}_{1}\rangle_{2},\\
 |\widetilde{\Psi}_{2}\rangle&\propto a|\widetilde{0}_{2}\rangle_{2}+e^{\frac{2\pi i}{3}}b|\widetilde{1}_{2}\rangle_{2},\\
|\widetilde{\Psi}_{3}\rangle&\propto a|\widetilde{0}_{0}\rangle_{2}-b|\widetilde{1}_{0}\rangle_{2},\\
 |\widetilde{\Psi}_{4}\rangle&\propto  a|\widetilde{0}_{1}\rangle_{2}-e^{\frac{\pi i}{3}}b |\widetilde{1}_{1}\rangle_{2},\\
 |\widetilde{\Psi}_{5}\rangle&\propto  a|\widetilde{0}_{2}\rangle_{2}-e^{\frac{2\pi i}{3}}b|\widetilde{1}_{2}\rangle_{2}.\\
\end{aligned}
\end{equation}
The final mixture for a logical qubit $|\bar{\Psi}\rangle=\frac{a|\bar{0}\rangle+b|\bar{1}\rangle}{\sqrt{1+2\operatorname{Re}(a^{*}b\langle \bar{0}|\bar{1}\rangle)}}$ can thus be written in the form
\begin{equation}
\label{eq: L2mixture}
\begin{aligned}
\bar{\rho}&=p_{0}\left(\frac{1+2\operatorname{Re}(a^{*}b\langle \widetilde{0}_{0}|\widetilde{1}_{0}\rangle)}{1+2\operatorname{Re}(a^{*}b\langle \bar{0}_{0}|\bar{1}_{0}\rangle)}\right)|\widetilde{\Psi}_{0}\rangle\langle \widetilde{\Psi}_{0}|,\\
&+p_{1}\left(\frac{1+2\operatorname{Re}(a^{*}be^{\frac{\pi i}{3}}\langle \widetilde{0}_{1}|\widetilde{1}_{1}\rangle)}{1+2\operatorname{Re}(a^{*}b\langle \bar{0}_{0}|\bar{1}_{0}\rangle)}\right)|\widetilde{\Psi}_{1}\rangle\langle \widetilde{\Psi}_{1}|,\\
&+p_{2}\left(\frac{1+2\operatorname{Re}(a^{*}be^{\frac{2\pi i}{3}}\langle \widetilde{0}_{2}|\widetilde{1}_{2}\rangle)}{1+2\operatorname{Re}(a^{*}b\langle \bar{0}_{0}|\bar{1}_{0}\rangle)}\right)|\widetilde{\Psi}_{2}\rangle\langle \widetilde{\Psi}_{2}|,\\
&+p_{3}\left(\frac{1-2\operatorname{Re}(a^{*}b\langle \widetilde{0}_{0}|\widetilde{1}_{0}\rangle)}{1+2\operatorname{Re}(a^{*}b\langle \bar{0}_{0}|\bar{1}_{0}\rangle)}\right)|\widetilde{\Psi}_{3}\rangle\langle \widetilde{\Psi}_{3}|,\\
&+p_{4}\left(\frac{1-2\operatorname{Re}(a^{*}be^{\frac{\pi i}{3}}\langle \widetilde{0}_{1}|\widetilde{1}_{1}\rangle}{1+2\operatorname{Re}(a^{*}b\langle \bar{0}_{0}|\bar{1}_{0}\rangle)}\right)|\widetilde{\Psi}_{4}\rangle\langle \widetilde{\Psi}_{4}|,\\
&+p_{5}\left(\frac{1-2\operatorname{Re}(a^{*}be^{\frac{2\pi i}{3}}\langle \widetilde{0}_{2}|\widetilde{1}_{2}\rangle)}{1+2\operatorname{Re}(a^{*}b\langle \bar{0}_{0}|\bar{1}_{0}\rangle)}\right)|\widetilde{\Psi}_{5}\rangle\langle \widetilde{\Psi}_{5}|,\\
 \end{aligned}
\end{equation}
where we also omitted the index indicating $L=2$ at the state vectors. For this code, the codeword probabilities $p_{i},~ i=0,...,5$, are given by 
\begin{equation}
p_{i}=\frac{1}{\sqrt{N}}\sum\limits_{m=0}^{\infty}\frac{(\alpha^{2}(1-\gamma))^{6m+i}}{(6m+i)!}.
 \end{equation}
From the mixed final state in Eq. \eqref{eq: L2mixture}, the non-deformation of the codewords becomes also manifest. Therefore, the KL-conditions
are approximately fulfilled. A recovery is therefore also approximately possible, provided the amplitudes are chosen large enough.
\newpage
\section{Amplitude restoration}
\label{sec: restauration}
 One effect of the realistic photon loss channel on the basic codewords is the damping of their amplitude. Consequently,
 the initial amplitude of the incoming logical state has to be restored.\\
 The first step in our quantum error correction process is a parity measurement that determines a certain error space. We have referred to this step as
 qubit recovery. For simplicity, the amplitude-damped codewords in the respective error space  are denoted as $|\widetilde{0}\rangle$ and $|\widetilde{1}\rangle$ in the following.
 The goal of amplitude restoration, the second step of our QEC, is to turn back the damped amplitudes $\sqrt{\gamma}\alpha$ to the initial amplitude
 for every codeword, $|\widetilde{0}\rangle\rightarrow |\bar{0}\rangle,~|\widetilde{1}\rangle\rightarrow |\bar{1}\rangle$. Later we will choose 
 to map the qubit with restored amplitudes from the error space back into the code space (where this step is not a necessity, but helpful w.r.t. 
 our one-way communication scheme).\\
 Our strategy is to teleport the damped qubit into a space spanned by undamped codewords using an encoded, asymmetric Bell state with one half
 damped and the other half undamped qubit (see Eq. \eqref{eq: asymBell}). In order to perform the Bell measurement onto a Bell basis expressed by non-orthogonal codewords, we 
 propose to first apply a probabilistic "filter operation" and then do a standard Bell measurement. Let us now describe this filter \cite{statedisc}.
 Since the codewords are not orthogonal for finite $\alpha$, they are not perfectly distinguishable. However, they can be written
 in some orthonormal basis $\{|x\rangle,|y\rangle\}$ as 
 \begin{equation}
 \begin{aligned}
 |\bar{0}\rangle&=b_{0}|x\rangle+b_{1}|y\rangle,\\
 |\bar{1}\rangle&=e^{i\phi}(b_{0}|x\rangle-b_{1}|y\rangle),
\end{aligned}
\end{equation}
where $b_{0}^{2}+b_{1}^{2}=1$ and $b_{0},b_{1}\in\mathbb{R}$ with $b_{0}>b_{1}$ without loss of generality. Furthermore, one has  $\langle \bar{0}|\bar{1}\rangle=e^{i\phi}(2b_{0}^{2}-1)$, such 
that
\begin{equation}
 \begin{aligned}
  b_{0}&=\sqrt{\frac{1+e^{-i\phi}\langle\bar{0}|\bar{1}\rangle}{2}},\\
  b_{1}&=\sqrt{\frac{1-e^{-i\phi}\langle\bar{0}|\bar{1}\rangle}{2}}.\\
  \end{aligned}
\end{equation}

Furthermore, we define the following  operators:  
 
 \begin{equation}
  A_{s}=\begin{pmatrix}
          \frac{b_{1}}{b_{0}}&0\\
          0&1
         \end{pmatrix},
~         
   A_{f}=\begin{pmatrix}
          \sqrt{1-\left(\frac{b_{1}}{b_{0}}\right)^{2}}&0\\
          0&0
         \end{pmatrix}.    
\end{equation}
  
As can easily be checked, one has $A_{s}^{\dagger}A_{s}+A_{f}^{\dagger}A_{f}=\mathbbm{1}$ and we refer to the non-unitary operations expressed
by $A_{s}$ and $A_{f}$ as a successful and a failed filter operation, respectively. A successful filter on the codewords leads to
\begin{equation}
\begin{aligned}
 A_{s}|\bar{0}\rangle&=b_{1}(|x\rangle+|y\rangle),\\
 A_{s}|\bar{1}\rangle&=e^{i\phi}b_{1}(|x\rangle-|y\rangle),
 \end{aligned}
 \end{equation}
 i.e. it maps the non-orthogonal codewords onto orthogonal states. Because the codewords cannot be perfectly distinguished, this cannot be done
 deterministically. In fact, the success probability for the filter operation is 
\begin{equation}
 P_{succ}=\langle \bar{0}|A_{s}^{\dagger}A_{s}|\bar{0}\rangle=\langle \bar{1}|A_{s}^{\dagger}A_{s}|\bar{1}\rangle=2-2b_{0}^{2}=1-|\langle\bar{0}|\bar{1}\rangle|. 
 \end{equation}
 
 Before proceeding, we illustrate the idea using the $L=0$ cat code, whose codewords are $|\bar{0}\rangle=|\alpha\rangle$ and 
 $|\bar{1}\rangle=|-\alpha\rangle$ with real overlap $\langle \alpha|-\alpha\rangle=e^{-2\alpha^{2}}$. The corresponding orthogonal
 basis is the cat state basis
 \begin{equation}
  \begin{aligned}
   |x\rangle=\frac{1}{\sqrt{N_{+}}}(|\alpha\rangle+|-\alpha\rangle),\\
   |y\rangle=\frac{1}{\sqrt{N_{-}}}(|\alpha\rangle-|-\alpha\rangle).\\
  \end{aligned}
\end{equation}
Since the overlap is real, we have $\phi=0$ and find
\begin{equation}
 \begin{aligned}
   b_{0}&=\sqrt{\frac{1+\exp(-2\alpha^{2})}{2}},\\
   b_{1}&=\sqrt{\frac{1-\exp(-2\alpha^{2})}{2}}.\\
 \end{aligned}
\end{equation}
\noindent
The probability for successfully distinguishing $|\bar{0}\rangle$ and  $|\bar{1}\rangle$ is therefore $P_{succ}=1-\exp(-2\alpha^{2})$ (so-called
unambiguous state discrimination).\\
For the teleportation-based amplitude restoration scheme, we need the Bell states in the (known) error space (note the normalization factor due to non-orthogonality of the codewords):
\begin{equation}
\begin{aligned}
 |\widetilde{\phi}_{+}\rangle=\frac{1}{\sqrt{N_{\widetilde{\phi}_{+}}}}\frac{|\widetilde{0}\rangle|\widetilde{0}\rangle+|\widetilde{1}\rangle|\widetilde{1}\rangle}{\sqrt{2}}\\
 |\widetilde{\phi}_{-}\rangle=\frac{1}{\sqrt{N_{\widetilde{\phi}_{-}}}}\frac{|\widetilde{0}\rangle|\widetilde{0}\rangle-|\widetilde{1}\rangle|\widetilde{1}\rangle}{\sqrt{2}}\\
 |\widetilde{\psi}_{+}\rangle=\frac{1}{\sqrt{N_{\widetilde{\psi}_{+}}}}\frac{|\widetilde{0}\rangle|\widetilde{1}\rangle+|\widetilde{1}\rangle|\widetilde{0}\rangle}{\sqrt{2}}\\
 |\widetilde{\psi}_{-}\rangle=\frac{1}{\sqrt{N_{\widetilde{\psi}_{-}}}}\frac{|\widetilde{0}\rangle|\widetilde{1}\rangle-|\widetilde{1}\rangle|\widetilde{0}\rangle}{\sqrt{2}}\\
 \end{aligned}
 \end{equation}
 For later use for the Bennett decomposition in the teleportation step, one rearranges the former equations into
 \begin{equation}
 \begin{aligned}
  |\widetilde{0}\rangle|\widetilde{0}\rangle&=\frac{1}{\sqrt{2}}(\sqrt{N_{\widetilde{\phi}_{+}}}|\widetilde{\phi}_{+}\rangle+\sqrt{N_{\widetilde{\phi}_{-}}}|\widetilde{\phi}_{-}\rangle),\\
 |\widetilde{1}\rangle|\widetilde{1}\rangle&=\frac{1}{\sqrt{2}}(\sqrt{N_{\widetilde{\phi}_{+}}}|\widetilde{\phi}_{+}\rangle-\sqrt{N_{\widetilde{\phi}_{-}}}|\widetilde{\phi}_{-}\rangle),\\
 |\widetilde{0}\rangle|\widetilde{1}\rangle&=\frac{1}{\sqrt{2}}(\sqrt{N_{\widetilde{\psi}_{+}}}|\widetilde{\psi}_{+}\rangle+\sqrt{N_{\widetilde{\psi}_{-}}}|\widetilde{\psi}_{-}\rangle),\\
 |\widetilde{1}\rangle|\widetilde{0}\rangle&=\frac{1}{\sqrt{2}}(\sqrt{N_{\widetilde{\psi}_{+}}}|\widetilde{\psi}_{+}\rangle-\sqrt{N_{\widetilde{\psi}_{-}}}|\widetilde{\psi}_{-}\rangle)\\
 \end{aligned}
\end{equation} 

The amplitude restoration works as follows: a encoded qubit is sent through the channel whose output is a mixed state. As pointed out in the main text, the first step in
the error correction procedure is the parity measurement which determines the corresponding error space, i.e. the input qubit of our amplitude restoration is of the form
$|\omega\rangle=\frac{c_{0}|\widetilde{0}\rangle+c_{1}|\widetilde{1}}{\sqrt{N_{\omega}}}$ with unknown coefficients $c_{0}$ and $c_{1}$. According to the result of the parity measurement, the
following state must be generated:
\begin{equation}
\label{eq: asymBell}
|\hat{\phi}^{+}\rangle=\frac{1}{\sqrt{N_{\hat{\phi}_{+}}}}\frac{|\widetilde{0}\rangle|\bar{0}\rangle+|\widetilde{1}\rangle|\bar{1}\rangle}{\sqrt{2}}.
 \end{equation}

In total, we then have
\begin{align*}
|\omega\rangle\otimes |\hat{\phi}^{+}\rangle&=\frac{1}{\sqrt{N_{\omega}N_{\hat{\phi}^{+}}}}\frac{1}{\sqrt{2}}(c_{0}|\widetilde{0}\rangle|\widetilde{0}\rangle|\bar{0}\rangle+ \numberthis
c_{0}|\widetilde{0}\rangle|\widetilde{1}\rangle|\bar{1}\rangle)+c_{1}|\widetilde{1}\rangle|\widetilde{0}\rangle|\bar{0}\rangle+c_{1}|\widetilde{1}\rangle|\widetilde{1}\rangle|\bar{1}\rangle)\\
&=\frac{1}{\sqrt{N_{\omega}N_{\hat{\phi}^{+}}}}\frac{1}{2}[c_{0}\sqrt{N_{\widetilde{\phi}_{+}}}|\widetilde{\phi}_{+}\rangle|\bar{0}\rangle+
\displaybreak
c_{0}\sqrt{N_{\widetilde{\phi}_{-}}}|\widetilde{\phi}_{-}\rangle|\bar{0}\rangle+c_{0}\sqrt{N_{\widetilde{\psi}_{+}}}|\widetilde{\psi}_{+}\rangle|\bar{1}\rangle+
c_{0}\sqrt{N_{\widetilde{\psi}_{-}}}|\widetilde{\psi}_{-}\rangle|\bar{1}\rangle\\
&+c_{1}\sqrt{N_{\widetilde{\psi}_{+}}}|\widetilde{\psi}_{+}\rangle|\bar{0}\rangle-c_{1}\sqrt{N_{\widetilde{\psi}_{-}}}|\widetilde{\phi}_{-}\rangle|\bar{0}\rangle]
+c_{1}\sqrt{N_{\widetilde{\phi}_{+}}}|\widetilde{\phi}_{+}\rangle|\bar{1}\rangle-c_{1}\sqrt{N_{\widetilde{\phi}_{-}}}|\widetilde{\phi}_{-}\rangle|\bar{0}\rangle\\
&=\frac{1}{\sqrt{N_{\omega}N_{\hat{\phi}^{+}}}}\frac{1}{2}(\sqrt{N_{\widetilde{\phi}_{+}}}|\widetilde{\phi}_{+}\rangle(c_{0}|\bar{0}\rangle+c_{1}|\bar{1}\rangle)+\sqrt{N_{\widetilde{\phi}_{-}}}|\widetilde{\phi}_{-}\rangle(c_{0}|\bar{0}\rangle-c_{1}|\bar{1}\rangle)\\
&+\sqrt{N_{\widetilde{\psi}_{-}}}|\widetilde{\psi}_{+}\rangle(c_{0}|\bar{1}\rangle+c_{1}|\bar{0}\rangle)+\sqrt{N_{\widetilde{\psi}_{-}}}|\widetilde{\psi}_{-}\rangle(c_{0}|\bar{1}\rangle-c_{1}|\bar{0}\rangle)\\
&=\frac{\sqrt{N_{\chi_{1}}}}{\sqrt{N_{\omega}N_{\hat{\phi}^{+}}}}\frac{1}{2}\sqrt{N_{\widetilde{\phi}^{+}}}|\widetilde{\phi}_{+}\rangle\left(\frac{c_{0}|\bar{0}\rangle+c_{1}|\bar{1}\rangle}{\sqrt{N_{\chi_{1}}}}\right)
+\frac{\sqrt{N_{\chi_{2}}}}{\sqrt{N_{\omega}N_{\hat{\phi}^{+}}}}\frac{1}{2}\sqrt{N_{\widetilde{\phi}^{-}}}|\widetilde{\phi}_{-}\rangle\left(\frac{c_{0}|\bar{0}\rangle-c_{1}|\bar{1}\rangle}{\sqrt{N_{\chi_{2}}}}\right)\\
&+\frac{\sqrt{N_{\chi_{3}}}}{\sqrt{N_{\omega}N_{\hat{\phi}^{+}}}}\frac{1}{2}\sqrt{N_{\widetilde{\psi}^{+}}}|\widetilde{\psi}_{+}\rangle\left(\frac{c_{0}|\bar{1}\rangle+c_{1}|\bar{0}\rangle}{\sqrt{N_{\chi_{3}}}}\right)
+\frac{\sqrt{N_{\chi_{3}}}}{\sqrt{N_{\omega}N_{\hat{\phi}^{+}}}}\frac{1}{2}\sqrt{N_{\widetilde{\psi}^{-}}}|\widetilde{\psi}_{-}\rangle\left(\frac{c_{0}|\bar{1}\rangle-c_{1}|\bar{1}\rangle}{\sqrt{N_{\chi_{4}}}}\right)\\
&=\frac{\sqrt{N_{\chi_{1}}}}{\sqrt{N_{\omega}N_{\hat{\phi}^{+}}}}\frac{1}{2}\sqrt{N_{\widetilde{\phi}^{+}}}\left(\frac{|\widetilde{0}\rangle|\widetilde{0}\rangle+|\widetilde{1}\rangle|\widetilde{1}\rangle}{\sqrt{2}}\right)\left(\frac{c_{0}|\bar{0}\rangle+c_{1}|\bar{1}\rangle}{\sqrt{N_{\chi_{1}}}}\right)\\
&+\frac{\sqrt{N_{\chi_{2}}}}{\sqrt{N_{\omega}N_{\hat{\phi}^{+}}}}\frac{1}{2}\sqrt{N_{\widetilde{\phi}^{-}}}\left(\frac{|\widetilde{0}\rangle|\widetilde{0}\rangle-|\widetilde{1}\rangle|\widetilde{1}\rangle}{\sqrt{2}}\rangle\right)\left(\frac{c_{0}|\bar{0}\rangle-c_{1}|\bar{1}\rangle}{\sqrt{N_{\chi_{2}}}}\right)\\
&+\frac{\sqrt{N_{\chi_{3}}}}{\sqrt{N_{\omega}N_{\hat{\phi}^{+}}}}\frac{1}{2}\sqrt{N_{\widetilde{\psi}^{+}}}\left(\frac{|\widetilde{0}\rangle|\widetilde{1}\rangle+|\widetilde{1}\rangle|\widetilde{0}\rangle}{\sqrt{2}}\right)\left(\frac{c_{0}|\bar{1}\rangle+c_{1}|\bar{0}\rangle}{\sqrt{N_{\chi_{3}}}}\right)\\
&+\frac{\sqrt{N_{\chi_{3}}}}{\sqrt{N_{\omega}N_{\hat{\phi}^{+}}}}\frac{1}{2}\sqrt{N_{\widetilde{\psi}^{-}}}\left(\frac{|\widetilde{0}\rangle|\widetilde{1}\rangle-|\widetilde{1}\rangle|\widetilde{0}\rangle}{\sqrt{2}}\right)\left(\frac{c_{0}|\bar{1}\rangle-c_{1}|\bar{0}\rangle}{\sqrt{N_{\chi_{4}}}}\right).
\end{align*}
Note that each of the four different output qubits requires a different normalisation factor which we denote as $N_{\chi_{i}}, i=1,2,3,4$.
Like in a usual teleportation scheme, we have a superposition of tensor products of four Bell states and four different output qubits. Because the codewords
$|\widetilde{0}\rangle$ and $|\widetilde{1}\rangle$ are not orthogonal, the Bell measurement in this basis can not be performed deterministically.
Therefore, we apply the filter operation on the first two modes individually
which leads, after an additional Hadamard-Gate in $\{|x\rangle, |y\rangle\}$, to
\begin{equation}
\label{eq: Hadamard}
\begin{aligned}
|\omega\rangle\otimes |\hat{\phi}^{+}\rangle\rightarrow&\frac{b_{1}^{2}\sqrt{N_{\chi_{1}}}\sqrt{N_{\widetilde{\phi}^{+}}}}{\sqrt{N_{\omega}N_{\hat{\phi}^{+}}}}\left(\frac{|x\rangle|x\rangle+e^{2i\phi}|y\rangle|y\rangle}{\sqrt{2}}\right)\left(\frac{c_{0}|\bar{0}\rangle+c_{1}|\bar{1}\rangle}{\sqrt{N_{\chi_{1}}}}\right)\\
&+\frac{b_{1}^{2}\sqrt{N_{\chi_{2}}}\sqrt{N_{\widetilde{\phi}^{-}}}}{\sqrt{N_{\omega}N_{\hat{\phi}^{+}}}}\left(\frac{|x\rangle|x\rangle-e^{2i\phi}|y\rangle|y\rangle}{\sqrt{2}}\rangle\right)\left(\frac{c_{0}|\bar{0}\rangle-c_{1}|\bar{1}\rangle}{\sqrt{N_{\chi_{2}}}}\right)\\
&+e^{i\phi}\frac{b_{1}^{2}\sqrt{N_{\chi_{3}}}\sqrt{N_{\widetilde{\psi}^{+}}}}{\sqrt{N_{\omega}N_{\hat{\phi}^{+}}}}\left(\frac{|x\rangle|y\rangle+|y\rangle|x\rangle}{\sqrt{2}}\right)\left(\frac{c_{0}|\bar{1}\rangle+c_{1}|\bar{0}\rangle}{\sqrt{N_{\chi_{3}}}}\right)\\
&+e^{i\phi}\frac{b_{1}^{2}\sqrt{N_{\chi_{3}}}\sqrt{N_{\widetilde{\psi}^{-}}}}{\sqrt{N_{\omega}N_{\hat{\phi}^{+}}}}\left(\frac{|x\rangle|y\rangle-|y\rangle|x\rangle}{\sqrt{2}}\right)\left(\frac{c_{0}|\bar{1}\rangle-c_{1}|\bar{1}\rangle}{\sqrt{N_{\chi_{4}}}}\right)=:|\nu\rangle.
\end{aligned}
\end{equation}
Since $|x\rangle$ and $|y\rangle$ are orthogonal, the Bell measurement can be performed. Because the filter operation is non-deterministic, the 
whole teleportations scheme has a non-unit success probability which corresponds to the norm of the state in Eq. \eqref{eq: Hadamard},
\begin{align*}
P_{succ}=\langle\nu|\nu\rangle&=\frac{b_{1}^{4}}{N_{\omega}N_{\hat{\phi}^{+}}}(N_{\chi_{1}}N_{\widetilde{\phi}^{+}}+N_{\chi_{2}}N_{\widetilde{\phi}^{-}}+N_{\chi_{3}}N_{\widetilde{\psi}^{+}}+N_{\chi_{4}}N_{\widetilde{\psi}^{-}})\numberthis\\
\displaybreak
&=\frac{(1-b_{0}^{2})^{2}}{N_{\omega}N_{\hat{\phi}^{+}}}(N_{\chi_{1}}N_{\widetilde{\phi}^{+}}+N_{\chi_{2}}N_{\widetilde{\phi}^{-}}+N_{\chi_{3}}N_{\widetilde{\psi}^{+}}+N_{\chi_{4}}N_{\widetilde{\psi}^{-}})\\
&=\frac{(1-e^{-i\phi}\langle \widetilde{0}|\widetilde{1}\rangle)^{2}}{4N_{\omega}N_{\hat{\phi}^{+}}}(N_{\chi_{1}}N_{\widetilde{\phi}^{+}}+N_{\chi_{2}}N_{\widetilde{\phi}^{-}}+N_{\chi_{3}}N_{\widetilde{\psi}^{+}}+N_{\chi_{4}}N_{\widetilde{\psi}^{-}})\\
&=\frac{(1-e^{-i\phi}\langle \widetilde{0}|\widetilde{1}\rangle)^{2}}{4N_{\omega}N_{\hat{\phi}^{+}}}(N_{\chi_{1}}N_{\widetilde{\phi}^{+}}+N_{\chi_{2}}N_{\widetilde{\phi}^{-}}+N_{\chi_{3}}N_{\widetilde{\psi}^{+}}+N_{\chi_{4}}N_{\widetilde{\psi}^{-}})\\
&=\frac{(1-e^{-i\phi}\langle \widetilde{0}|\widetilde{1}\rangle)^{2}}{4(1+2\operatorname{Re}(c_{0}^{*}c_{1}\langle \widetilde{0}|\widetilde{1}\rangle))(1+\operatorname{Re}(\langle \widetilde{0}|\widetilde{1}\rangle\langle \bar{0}|\bar{1}\rangle))}(N_{\chi_{1}}N_{\widetilde{\phi}^{+}}+N_{\chi_{2}}N_{\widetilde{\phi}^{-}}+N_{\chi_{3}}N_{\widetilde{\psi}^{+}}+N_{\chi_{4}}N_{\widetilde{\psi}^{-}}).
\end{align*}
For an $L$-encoded qubit with coefficients $a$ and $b$, the total success probability for the one-way scheme is (see Sec. \ref{sec: oneway communication})
\begin{equation}
 P_{ow}=\left(\sum\limits_{k=0}^{2L+1}\widetilde{p}_{k}(a,b)\cdot P_{succ}\left[a|\widetilde{0}_{k}\rangle_{L}+\exp\left(\frac{k \pi i}{L+1}\right)b|\widetilde{1}_{k}\rangle_{L}\right]\right)^{\mathcal{L}/d_{0}},
\end{equation}
where $\mathcal{L}$ is the total distance, $d_{0}$ is the regular interval at which AR is performed, and the index "$k$" in the codewords is to be understood
as modulo $L+1$ to obtain the corresponding error space codewords (recall $q=0,...,L$). Note that the
sum goes over all components in the incoming mixed state because this probability does not correspond to the success probability of QEC but 
to the probability of the filter to succeed.

\end{document}